\documentclass[twocolumn,tighten,times,usenatbib,twocolappendix,numberedappendix]{aastex631}

\usepackage{amsmath}
\usepackage{float}
\usepackage{color}
\usepackage{url}
\usepackage{stmaryrd}

\usepackage[percent]{overpic}

\usepackage{etoolbox}

\usepackage{amsmath}
\usepackage{amssymb}
\usepackage{graphicx}
\usepackage[flushleft]{threeparttable}
\usepackage{placeins}
\usepackage{xcolor}

\usepackage{siunitx}

\hypersetup{breaklinks=true,colorlinks=true,linkcolor=blue,citecolor=blue,filecolor=blue,urlcolor=blue}
\makeatletter
\patchcmd{\NAT@citex}
  {\@citea\NAT@hyper@{%
     \NAT@nmfmt{\NAT@nm}%
     \hyper@natlinkbreak{\NAT@aysep\NAT@spacechar}{\@citeb\@extra@b@citeb}%
     \NAT@date}}
  {\@citea\NAT@nmfmt{\NAT@nm}%
   \NAT@aysep\NAT@spacechar\NAT@hyper@{\NAT@date}}{}{}

\patchcmd{\NAT@citex}
  {\@citea\NAT@hyper@{%
     \NAT@nmfmt{\NAT@nm}%
     \hyper@natlinkbreak{\NAT@spacechar\NAT@@open\if*#1*\else#1\NAT@spacechar\fi}%
       {\@citeb\@extra@b@citeb}%
     \NAT@date}}
  {\@citea\NAT@nmfmt{\NAT@nm}%
   \NAT@spacechar\NAT@@open\if*#1*\else#1\NAT@spacechar\fi\NAT@hyper@{\NAT@date}}
  {}{}

\usepackage{array}
\newcolumntype{C}[1]{>{\centering\let\newline\\\arraybackslash\hspace{0pt}}m{#1}}

\usepackage{xspace}

\def\araa{ARA\&A}
\def\apj{ApJ}
\def\apjl{ApJ}
\def\apjs{ApJS}

\def\apss{Ap\&SS}
\def\aap{A\&A}

\def\mnras{MNRAS}

\def\pasa{Publ.~Astron.~Soc.~Australia}
\def\pasp{PASP}
\def\pasj{PASJ}

\definecolor{burgundy}{rgb}{0.5, 0.0, 0.13}

\usepackage{xspace}
\usepackage{wasysym}

\newcommand{\orcidicon}{\includegraphics[width=0.26cm]{orcid-ID.eps}}

\newcommand{\orcidauthor}[1]{\href{https://orcid.org/#1}{\orcidicon}}

\shorttitle{Wind accretion in Symbiotic Systems}
\shortauthors{R. F. Maldonado et al.}

\makeatletter
\patchcmd{\frontmatter@RRAP@format}{(}{}{}{}
\patchcmd{\frontmatter@RRAP@format}{)}{}{}{}
\renewcommand\Dated@name{}
\makeatother

\begin{document}

\title{\large The impact of wind accretion in Evolving Symbiotic Systems}

\correspondingauthor{Raul F. Maldonado}
\email{r.maldonado@irya.unam.mx}

\author[0000-0002-2236-7554]{Raul F. Maldonado}
\affiliation{Instituto de Radioastronom\'{i}a y Astrof\'{i}sica, Universidad Nacional Aut\'{o}noma de M\'{e}xico, Morelia 58089, Mich., Mexico}

\author[0000-0002-5406-0813]{Jes\'{u}s~A.~Toal\'{a}}
\affiliation{Instituto de Radioastronom\'{i}a y Astrof\'{i}sica, Universidad Nacional Aut\'{o}noma de M\'{e}xico, Morelia 58089, Mich., Mexico}

\author[0000-0002-0616-8336]{Janis B. Rodr\'{i}guez-Gonz\'{a}lez}
\affiliation{Instituto de Radioastronom\'{i}a y Astrof\'{i}sica, Universidad Nacional Aut\'{o}noma de M\'{e}xico, Morelia 58089, Mich., Mexico}

\author[0000-0001-9936-6165]{Emilio Tejeda}
\affiliation{SECIHTI - Instituto de F\'{i}sica y Matem\'{a}ticas, Universidad Michoacana de San Nicol\'{a}s de Hidalgo, Ciudad Universitaria, 58040 Morelia, Mich., Mexico}

\date[]{Accepted to ApJ}

\begin{abstract}
\noindent 
We investigate the impact of geometric corrections to the Bondi-Hoyle-Lyttleton (BHL) accretion scheme applied to evolving symbiotic systems. We model systems where 0.7 and 1 M$_\odot$ white dwarfs accrete material from Solar-like stars with initial masses of 1, 2, and 3 M$_\odot$. The primary star is evolved using the MESA stellar evolution code, while the orbital dynamics of the system are calculated using REBOUND. The analysis focuses on systems evolving through the red giant branch and the thermally-pulsating asymptotic giant branch phases that do not experience a Wind Roche Lobe Overflow phase. We compare three scenarios: no accretion, standard BHL accretion, and the improved wind accretion. The choice of accretion prescription critically influences the evolution of symbiotic systems. Simulations using the modified model did not reach the Chandrasekhar limit, suggesting that type Ia supernova progenitors require accretors originating from ultra-massive WDs. In contrast, the standard BHL model predicts WD growth to this limit in compact systems. This discrepancy suggests that population synthesis studies adopting the traditional BHL approach may yield inaccurate results. The revised model successfully reproduces the accretion properties of observed symbiotic systems and predicts transitions between different accretion regimes driven by donor mass-loss variability. These results emphasize the need for updated wind accretion models to accurately describe the evolution of symbiotic binaries.
\end{abstract}

\keywords{
\href{http://astrothesaurus.org/uat/154}{Binary Stars~(154)}; 
\href{http://astrothesaurus.org/uat/1578}{Stellar accretion~(1578)};
\href{http://astrothesaurus.org/uat/1636}{Stellar Winds~(1636)};
\href{http://astrothesaurus.org/uat/1674}{Symbiotic binary stars~(1674)}
}

\section{Introduction}
\label{sec:intro}

Symbiotic stars are binary systems in which a white dwarf (WD) accretes material from its late-type companion, producing a variety of electromagnetic signatures \citep[see][]{Merc2025}. Studying accretion in symbiotic systems is crucial, because they are very likely the progenitors of unique objects, such as Barium stars \citep{Bidelman1951} and carbon- and $s$-element-enhanced metal-poor stars \citep{Beers2005}. Besides cataclysmic variables, symbiotic systems also produce nova-like events and are potential progenitors of type Ia supernovae with cosmological implications \citep[see][]{Mukai2017,Chomiuk2021}.

The accretion process in symbiotic systems can be studied through various types of observations, such as the rotation of accretion discs \citep[e.g.,][]{Leedjarv1994,Robinson1994,Zamanov2024}, associated X-ray emission \citep[e.g.,][]{Toala2023,VT2024,Zhekov2019}, or the flickering effect observed mainly in optical wavelengths \citep[e.g.,][]{Sokoloski2001,Gromadzki2006,Merc2024}.

It has been widely accepted that accretion through wind capture in a Bondi-Hoyle-Lyttleton (BHL)-like scenario \citep{Hoyle1939,Bondi1944,Bondi1952} is not applicable to symbiotic systems, particularly when the velocity of the stellar wind ($v_\mathrm{w}$) of the mass-losing, late-type star is lower than the orbital velocity ($v_\mathrm{o}$) of the accreting WD \citep[e.g.,][]{Boffin2015,Hansen2016}. For $v_\mathrm{w}<v_\mathrm{o}$, the standard implementation of the BHL formalism in binary stars overestimates the mass accretion rate, predicting in some cases a non-physical value exceeding unity. This challenge has been discussed in several numerical simulations \citep[see, e.g.,][]{HE2013,Malfait2024,Saladino2019,Theuns1996}.

An alternative accretion mechanism known as the Wind-Roche Lobe Overflow (WRLO) has been proposed to study systems in which $v_\mathrm{w} < v_\mathrm{o}$ \citep{Mohamed2007}. This model proposes that if the wind injection radius of the mass donor star is similar or larger than the Roche Lobe radius ($R_\mathrm{RL}$) matter is transferred onto the companion through the first Lagrangian point. \citet[][]{Mohamed2012} used the dust condensation radius ($R_\mathrm{cond}$) as a proxy of the wind injection radius in late-type stars, following from the idea that their winds are dust-driven. Their simulations showed that if $R_\mathrm{cond} < R_\mathrm{RL}$, the wind capture accretion regime is recovered.

However, the problem has persisted since theoretical works still turn to BHL-like prescriptions to benchmark the accretion of WDs \citep[see, e.g.,][]{Chen2018,Vathachira2024}. Several groups tried to alleviate the mass accretion efficiency problem by including arbitrary efficiency factors to the BHL formalism \citep[e.g.,][]{Nagae2004,Malfait2024,Richards2024,SaladinoPols2019,Li2023} or cut-off values \citep[e.g.,][]{Saladino2019}, and those recipes are included in population synthesis studies \citep[see, e.g.,][]{Hurley2002,Izzard2009,Osborn2025,Andrews2024}.

\citet{TejedaToala2025} (hereinafter Paper I) have recently presented analytical modifications to the standard BHL formalism applied to accreting binaries that resulted in more consistent mass accretion efficiencies when compared with numerical simulations in the $v_\mathrm{w} < v_\mathrm{o}$ regime. Paper I showed that for circular orbits, the mass accretion efficiency $\eta$ can be expressed simply by two dimensionless quantities, the mass ratio $q=m_2/(m_1+m_2)$ and that corresponding to the velocity $w=v_\mathrm{w}/v_\mathrm{o}$, where $m_1$ and $m_2$ are the masses of the donor star and the accreting WD, respectively. Moreover, their approach allowed them to peer into the properties of accreting objects in eccentric orbits, resulting in intricate and more complex evolution of such systems.

In this paper, we build on the model proposed in Paper I to investigate the impact of accretion processes in evolving symbiotic systems. Specifically, we examine the detailed evolution of a WD in circular orbit around mass-losing, Solar-like stars with initial masses of 1, 2, and 3 M$_\odot$. A direct comparison with known symbiotic systems is also presented. We give some predictions for the final destination of symbiotic systems as possible type Ia supernova. Note that an extension of this work but for the case of eccentric orbits and systems entering a WRLO phase will be presented in subsequent papers.

This paper is organized as follows. In Section~\ref{sec:methods} we present our methodology, which includes the combination of stellar evolution models with dynamical simulations. In Section~\ref{sec:results} we present our results and a discussion is presented in Section~\ref{sec:discussion}. A summary is finally presented in Section~\ref{sec:summary}.

\section{Methods}
\label{sec:methods}

\subsection{Stellar Evolution Models}
\label{sec:MESA}

Evolution models for the donor star were generated using the Modules for Experiments of Stellar Evolution (MESA) code \citep{Paxton2011}.  Numerous improvements have been implemented since its initial release, and further details can be found in the associated papers \citep[see][]{Paxton2013,Paxton2015,Paxton2018,Paxton2019,Jermyn2023}. The r15140 version of MESA \citep{Paxton2019} is used to create our models\footnote{The MESA input files used in this study are available via Zenodo \dataset[10.5281/zenodo.15652665]{https://doi.org/10.5281/zenodo.15652665}}.

This study focuses on non-rotating stellar evolution models for Solar-like stars with initial masses of 1, 2, and 3 M$_\odot$. The calculations encompass the evolutionary phases from the main sequence to the WD stage. The initial metallicity was set to $Z=0.02$, and no magnetic field was included in the simulations. The mass-loss rate ($\dot{M}_\mathrm{w}$) during the red giant branch (RGB) phase was determined using the standard wind efficiency of 0.5 from \citet{Reimers1975} while, for the asymptotic giant branch (AGB) phase, the mass-loss rate from \citet{Bloecker1995} was adopted with a wind efficiency of 0.1.

\begin{figure}
\begin{center}
\includegraphics[width=\linewidth]{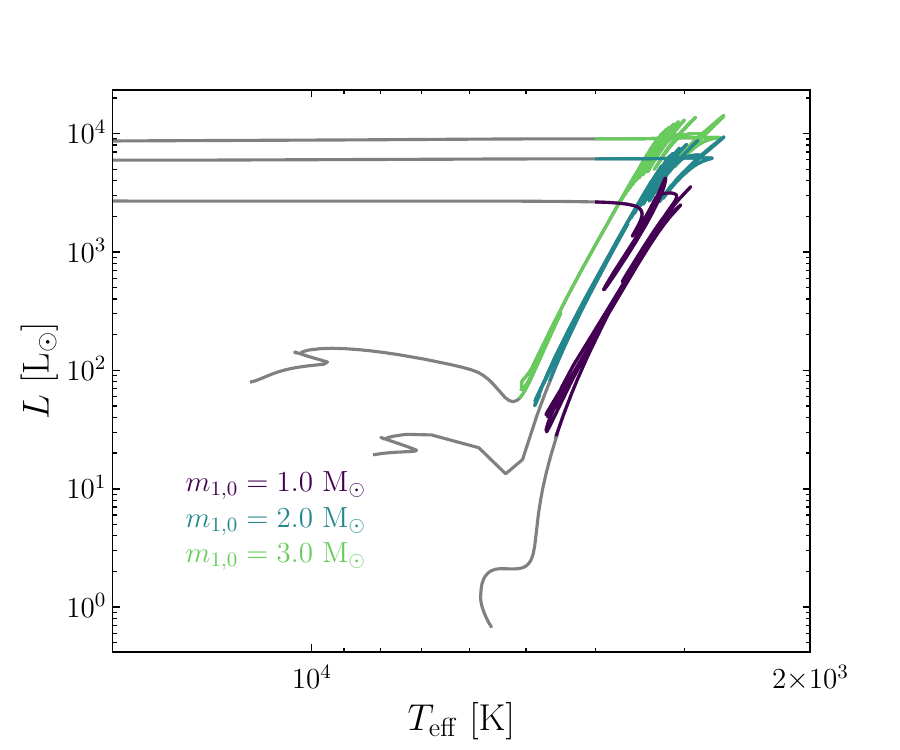}
\end{center}
\caption{Evolutionary tracks of the stellar evolution models used in this paper. The coloured regions depict the evolution period when the REBOUND integrations are preformed, from the ascending RGB phase to the end of the TPAGB phase, log$_{10}(T_\mathrm{eff}/\mathrm{K})=3.6$.}
\label{fig:hr_diag}
\end{figure}

\begin{figure*}
\begin{center}
\includegraphics[width=0.7\linewidth]{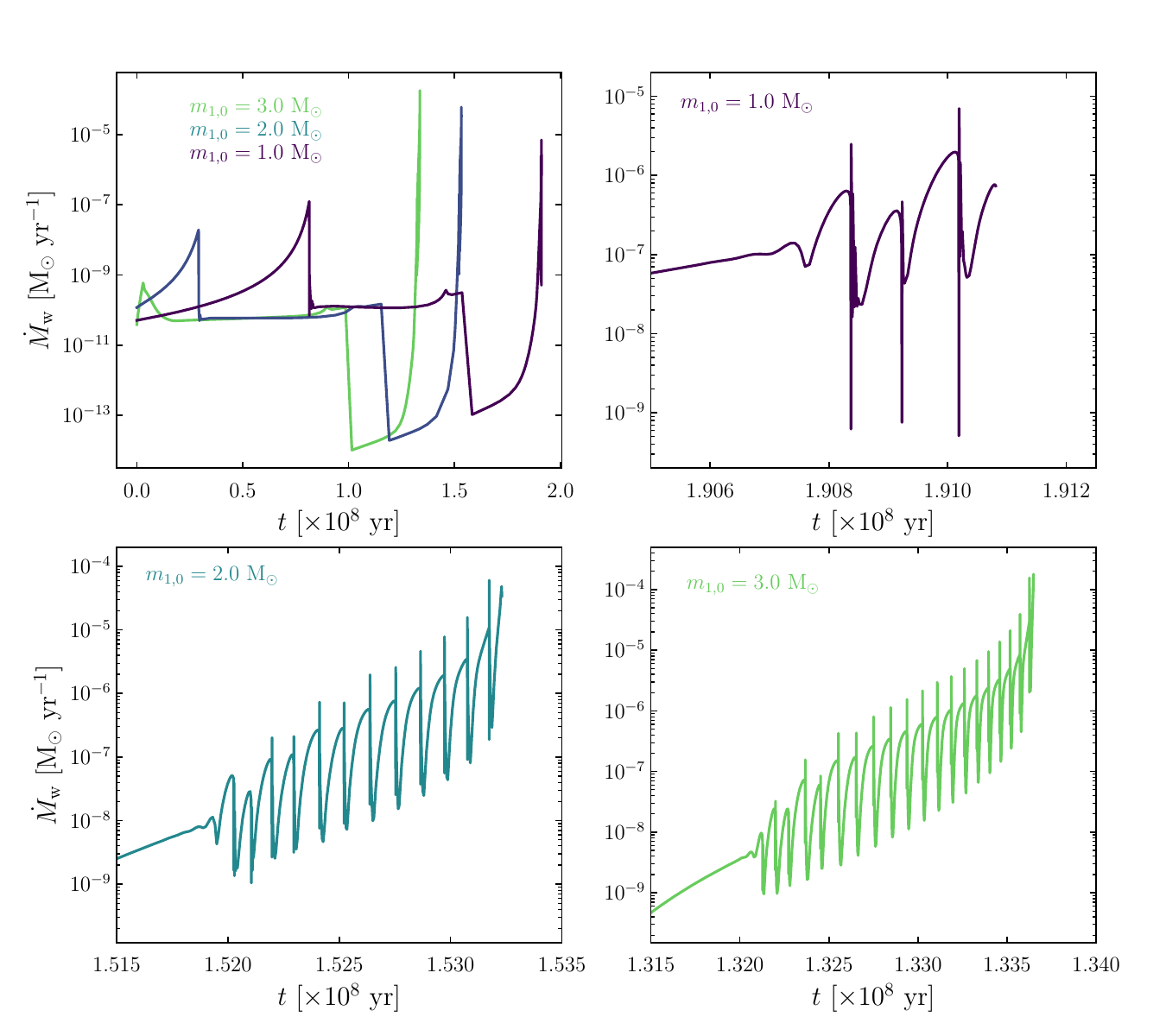}
\end{center}
\caption{Mass-loss rate $\dot{M}_\mathrm{w}$ as a function of time for the stellar evolution models used in this work. The upper left panel show the complete wind evolution from the RGB to the TPAGB phases while the other panels present details of the TPAGB phases of the different models.}
\label{fig:mdot}
\end{figure*}

Fig.~\ref{fig:hr_diag} illustrates the evolutionary tracks of the three stellar models in the Hertzsprung-Russell diagram, with the interval between the onset of the RGB and the end of the thermally pulsating asymptotic giant branch (TPAGB) phases highlighted in colour. This interval is characterized by significant mass loss, as illustrated in Fig.~\ref{fig:mdot}.

\begin{table}
\begin{center}
\caption{Details of the stellar evolution models used in this work. $m_{1,0}$ denotes the mass of the evolving star at the ZAMS, $m_\mathrm{1,iRGB}$ the mass at the beginning of the RGB, $m_\mathrm{1,fAGB}$ the mass at the end of the TPAGB, $\Delta m_1$ the total mass lost, $t_\mathrm{iRGB}$ the time at the beginning og the RGB phase, $t_\mathrm{fAGB}$ the time at the end of the TPAGB, and $\Delta t$ is the total time of the calculations. See Section~\ref{sec:MESA}.}
%\footnotesize
\setlength{\tabcolsep}{0.6\tabcolsep}  
%\scriptsize
\begin{tabular}{lcccc}
\hline
$m_\mathrm{1,0}$ & [M$_\odot$] & 1.0  & 2.0 & 3.0  \\
$m_\mathrm{1,iRGB}$   & [M$_\odot$] & 0.992 & 1.992 & 2.995  \\
$m_\mathrm{1,fAGB}$   & [M$_\odot$] & 0.534 & 0.586 & 0.644  \\
$\Delta m_1$ & [M$_\odot$] & 0.458 & 1.406 & 2.351  \\
\hline
$t_\mathrm{iRGB}$   & [Myr]   & 12142.716 & 1047.154  & 317.163  \\
$t_\mathrm{fAGB}$   & [Myr]   & 12333.797 & 1200.385 & 450.812 \\
$t_\mathrm{TOT}$    & [Myr]   & 191.081 & 153.230 & 133.649 \\
%\hline
\hline
\end{tabular}
\label{tab:models}
\end{center}
\end{table}

Table~\ref{tab:models} summarizes key parameters of the stellar evolution models: {\it i}) initial mass at the zero-age main sequence ($m_\mathrm{1,0}$), {\it ii}) mass at the beginning of the RGB phase ($m_\mathrm{1,iRGB}$), {\it iii}) mass at the end of the TPAGB ($m_\mathrm{1,fAGB}$), {\it iv}) total mass lost between the RGB and TPAGB phases ($\Delta m_1 = m_\mathrm{1,iRGB} - m_\mathrm{1,fAGB}$), {\it v}) age at the beginning of the RGB phase ($t_\mathrm{iRGB}$), {\it vi}) age at the end of the TPAGB phase ($t_\mathrm{fAGB}$), and {\it vii}) time duration between these two phases ($t_\mathrm{TOT} = t_\mathrm{fAGB} - t_\mathrm{iRGB}$).

\begin{figure*}
\begin{center}
\includegraphics[width=0.7\linewidth]{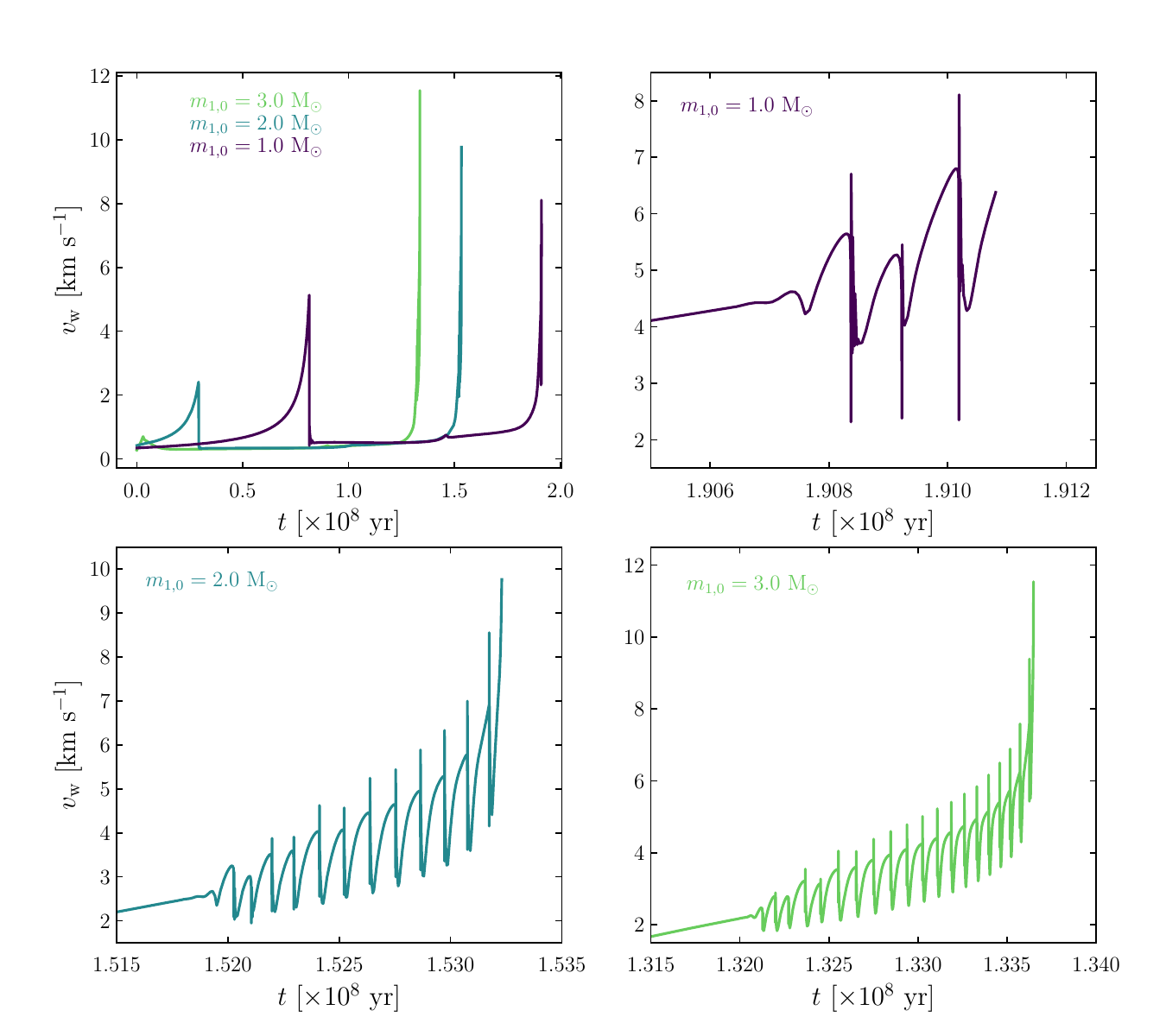}
\end{center}
\caption{Stellar wind velocity $v_\mathrm{w}$ as a function of time for the stellar evolution models used in this work. The upper left panel show the complete wind evolution during the RGB and TPAGB phases while the other panels present details of the TPAGB phases of the different models.}
\label{fig:vel_wind}
\end{figure*}

The terminal velocity of the stellar wind was calculated using the empirical relation derived by \citet{Verbena2011}:
\begin{equation}
    v_\mathrm{w} = 0.05 \left( \frac{L_1}{\mathrm{L}_\odot} \cdot \frac{\mathrm{M}_\odot}{m_1}\right)^{0.57} {\mathrm{km}~\mathrm{s}^{-1}},
\label{eq:verbena}
\end{equation}
\noindent where $L_1$ is the luminosity of the evolving star with mass $m_1$. Fig.~\ref{fig:vel_wind} shows the results of applying Eq.~(\ref{eq:verbena}) to the stellar evolution models obtained with MESA.

It is important to note that the wind velocity generally varies with distance from the stellar surface due to the acceleration of dust grains by radiation. However, given the relatively large orbital separations in the simulations discussed below, we can safely assume that the wind has reached its terminal velocity by the time it interacts with the accreting WD.

\subsection{Binary evolution}

The orbital evolution of each symbiotic system was calculated using the extensively-tested N-body package REBOUND \citep{Rein2012}, employing the 15th-order, implicit integrator with adaptive step-size control \citep[IAS15;][]{Rein2015}.

All simulations incorporate the effects of mass loss for the donor star $m_1$ due to its stellar wind. To account for the WD's response to accretion, we considered three scenarios:
\begin{enumerate}
    \item No accretion ($m_2$ remains constant), $\dot{M}_\mathrm{acc}=0$.
    \item Modified accretion model presented in Paper I, with $\dot{M}_\mathrm{acc}$=$\eta \dot{M}_\mathrm{w}$, where
\begin{equation}
\eta = \left(\frac{q}{1+w^2}\right)^2.
\label{eq:TejedaToala}
\end{equation}    
    \item Standard implementation of the BHL model, with $\dot{M}_\mathrm{acc}$=$\eta_\mathrm{BHL} \dot{M}_\mathrm{w}$, where
\begin{equation} 
\eta_\mathrm{BHL} = \frac{q^2}{w (1 + w^2)^{3/2}}. 
\label{eq:BHL}
\end{equation}    
\end{enumerate}
Remembering that $q$ represents the dimensionless mass ratio and $w$ represents the dimensionless velocity ratio, defined as
\begin{gather}
    q =\frac{m_2}{m_1 + m_2} ,
\label{eq:q}\\
w = \frac{v_\mathrm{w}}{v_\mathrm{o}},
\label{eq:w}
\end{gather}
where
\begin{equation}
    v_\mathrm{o} = \sqrt{\frac{G (m_1 + m_2)}{a}}
\end{equation}
is the orbital velocity, $a$ is the instantaneous orbital separation, and $G$ is the gravitational constant. 

A comparison of the accretion prescriptions given in Eqs.~\eqref{eq:TejedaToala} and \eqref{eq:BHL} is crucial, as the standard BHL model is widely used in the literature for evolving symbiotic systems \citep[see, e.g.,][]{Liu2017,Saladino2019,Vathachira2024}.

As noted in Section \ref{sec:intro}, the standard BHL model can predict non-physical accretion efficiencies greater than 1 for slow wind speeds ($w < 1$). To prevent this, we capped the BHL accretion efficiency at $\eta_\mathrm{BHL}=1$ whenever Eq.~(\ref{eq:BHL}) yielded values greater than unity.

The simulations were conducted with two different initial masses for the accreting WD: $m_{2,0}=0.7$ and 1.0 M$_\odot$. This choice stems from the assumption that the WD's progenitor was initially more massive than the donor star in the symbiotic system.

Because the donor star's mass-loss rate and wind velocity change over time, both $m_1$ and $m_2$ vary throughout the simulation. Consequently, parameters such as orbital separation ($a$), orbital velocity ($v_\mathrm{o}$), velocity ratio ($w$), mass ratio ($q$), accretion efficiency ($\eta$), and accretion rate ($\dot{M}_\mathrm{acc}$) are all functions of time ($t$).

\begin{figure*}
\begin{center}
\includegraphics[width=0.92\linewidth]{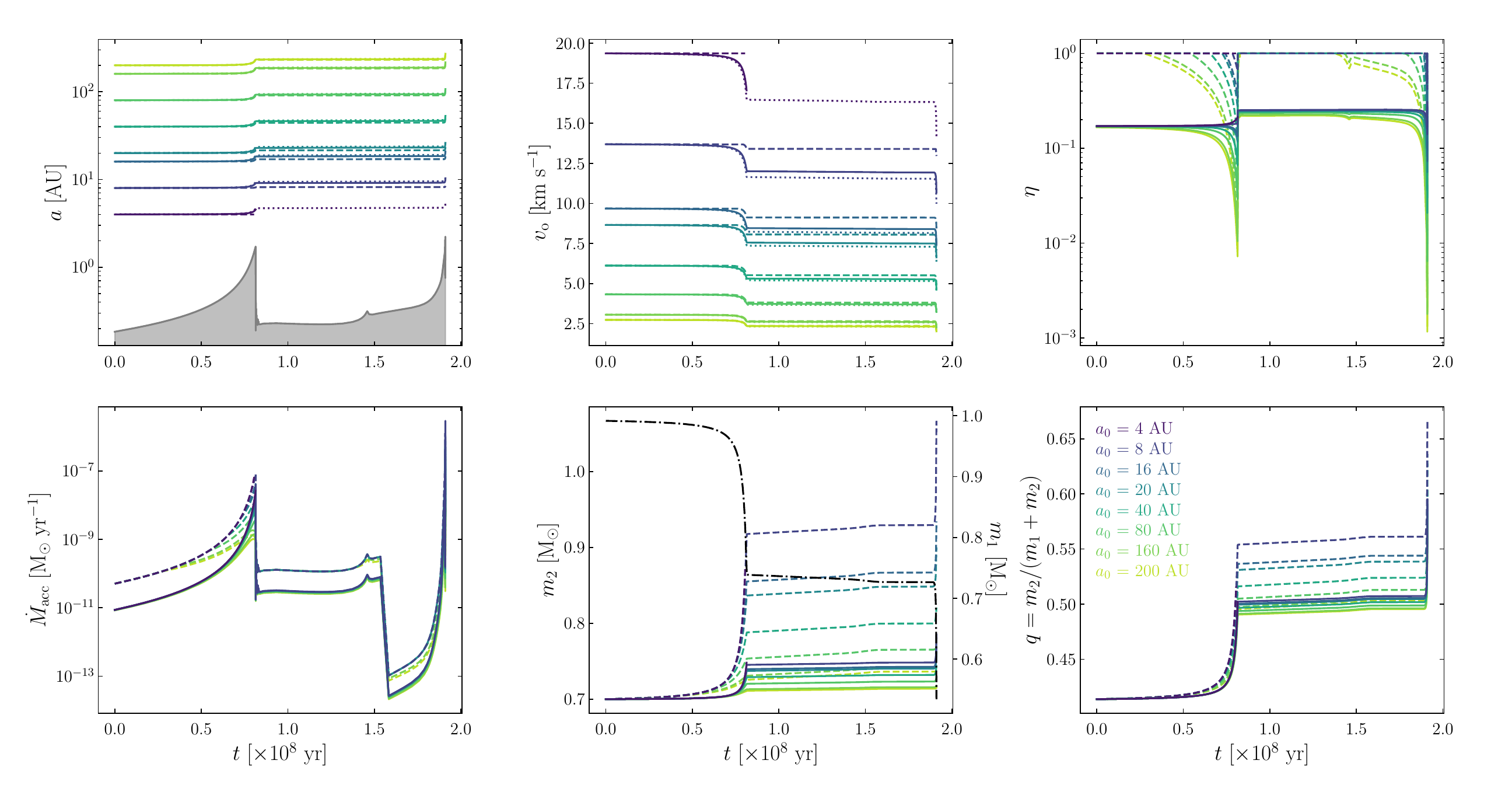}
\end{center}
\caption{Evolution with time of the orbital separation $a$ (top left), orbital velocity $v_\mathrm{o}$ (top middle), mass accretion efficiency $\eta$ (top right), mass accretion rate $\dot{M}_\mathrm{acc}=\eta\dot{M_\mathrm{w}}$ (bottom left), the WD mass $m_2$ (bottom middle) and $q=m_2/(m_1+m_2)$ (bottom right) for binary systems with initial stellar components with masses of $m_{1,0}=1.0$ M$_\odot$ and $m_{2,0}=0.7$~M$_\odot$. Solid lines show the results from simulations with accretion adopting the modified wind accretion model presented in Paper I, dashed lines show the results of simulations with the classical BHL accretion formalism and, dotted lines  are simulations without accretion. The shaded gray region in the top left panel depicts the evolution of the dust condensation radius $R_\mathrm{cond}$. The bottom middle panel also shows the evolution of the mass of the primary star $m_1$ with (black) dash-dotted line.}
\label{fig:time_evol_1M}
\end{figure*}
\begin{figure*}
\begin{center}
\includegraphics[width=0.92\linewidth]{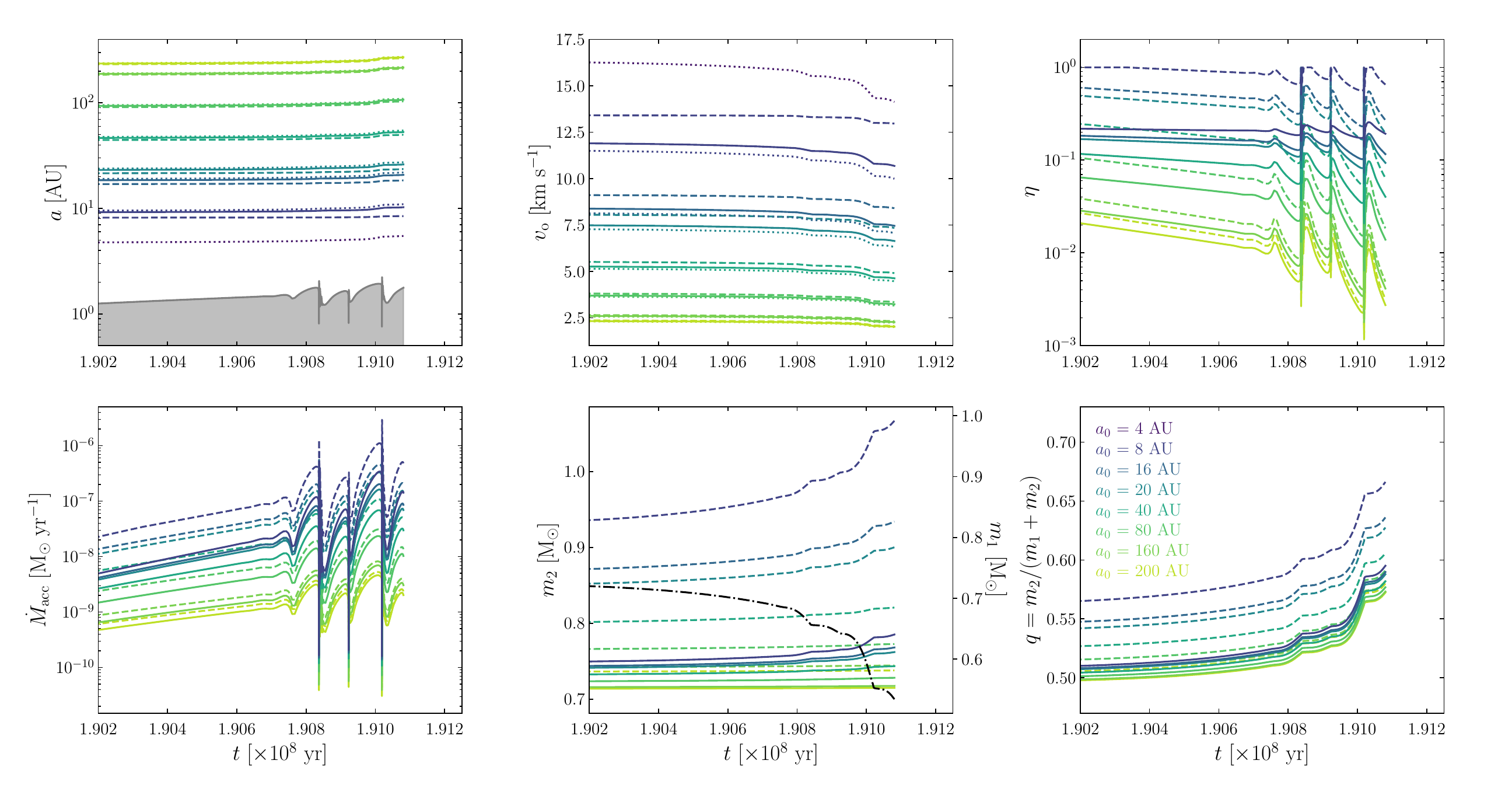}
\end{center}
\caption{Same as Fig.~\ref{fig:time_evol_1M} but only for times around the TPAGB phase.}
\label{fig:time_evol_1M_zoom}
\end{figure*}

Each simulation spanned the evolution of the binary system from the onset of the RGB phase past the end of the TPAGB phase, specifically until the effective temperature of the donor star reached $\log_{10}(T/\mathrm{K}) = 3.6$ after the TPAGB phase (see Fig.~\ref{fig:hr_diag}). This ensures that our simulated systems include D-type and S-type symbiotic binaries which are accepted to be associated with RGB and AGB stars, respectively \citep[see][and references therein]{Akras2019}.

The simulations were terminated if the WD accreted enough mass to reach the Chandrasekhar limit (1.4 M$_\odot$), at which point it would theoretically explode as a Type Ia supernova (a phenomenon not modelled in this study).

We also want to avoid our systems to enter the WRLO phase. Testing this alternative accretion mechanisms in combination with evolving symbiotic systems is out of the scope of the present work and will be pursuit in a subsequent paper. We adopted the condition proposed by \citet{Mohamed2012} where the dust condensation radius $R_\mathrm{cond}$ should be smaller than the Roche Lobe effective radius $R_\mathrm{RL}$ to ensure the systems to evolve purely through a wind capture accretion regime. If any of the modelled systems reach this condition, the simulation is stopped. The details on the calculation of $R_\mathrm{cond}$ and $R_\mathrm{RL}$ are presented in Appendix~\ref{app:WRLO}. We remark that such condition only affected the most compact systems ($a_0 = 4$ AU).

\section{Results}
\label{sec:results}

In this section we report the evolution of symbiotic systems across a range of initial conditions and accretion models. We considered primary star masses initially set to $m_{1,0} = 1$, 2, and 3 M$_\odot$, accreting WD masses initially set to $m_{2,0} = 0.7$ and 1.0 M$_\odot$, and initial binary separations of $a_0=4$, 8, 16, 20, 40, 80, 160, and 200 AU. This resulted in 48 unique initial configurations. For each configuration we conducted three types of dynamical simulations: {\it i}) no accretion, {\it ii}) accretion using the modified model described in Paper I, and {\it iii}) accretion using the standard BHL model, leading to a total of 144 simulations. 

\begin{figure*}
\begin{center}
\includegraphics[width=0.92\linewidth]{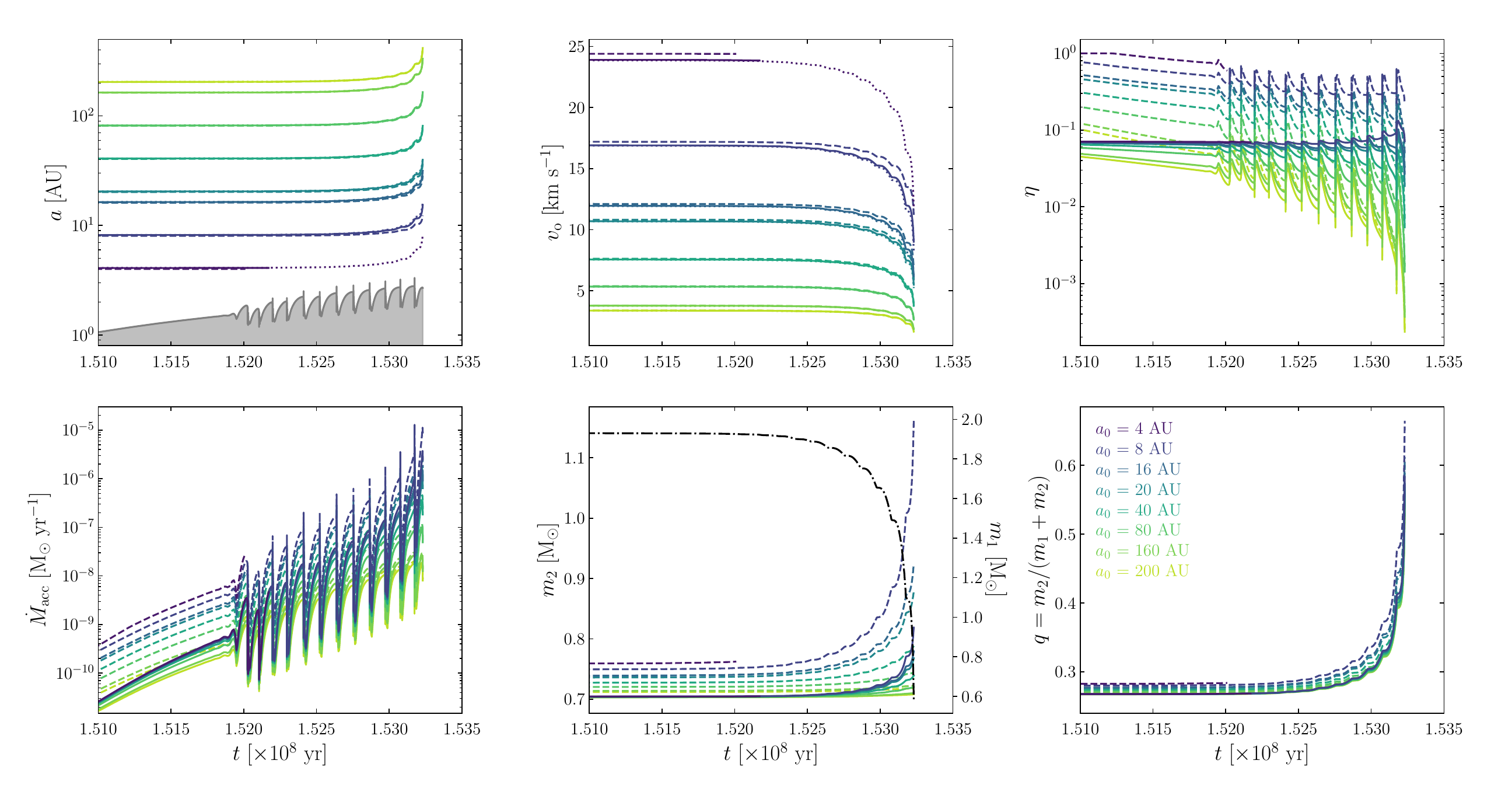}
\end{center}
\caption{The same as Fig.~\ref{fig:time_evol_1M_zoom} but for models with initial configuration of $m_{1,0}=2$ M$_\odot$ and $m_{2,0}=0.7$ M$_\odot$.}
\label{fig:time_evol_2M_zoom}
\end{figure*}

\begin{figure*}
\begin{center}
\includegraphics[width=0.92\linewidth]{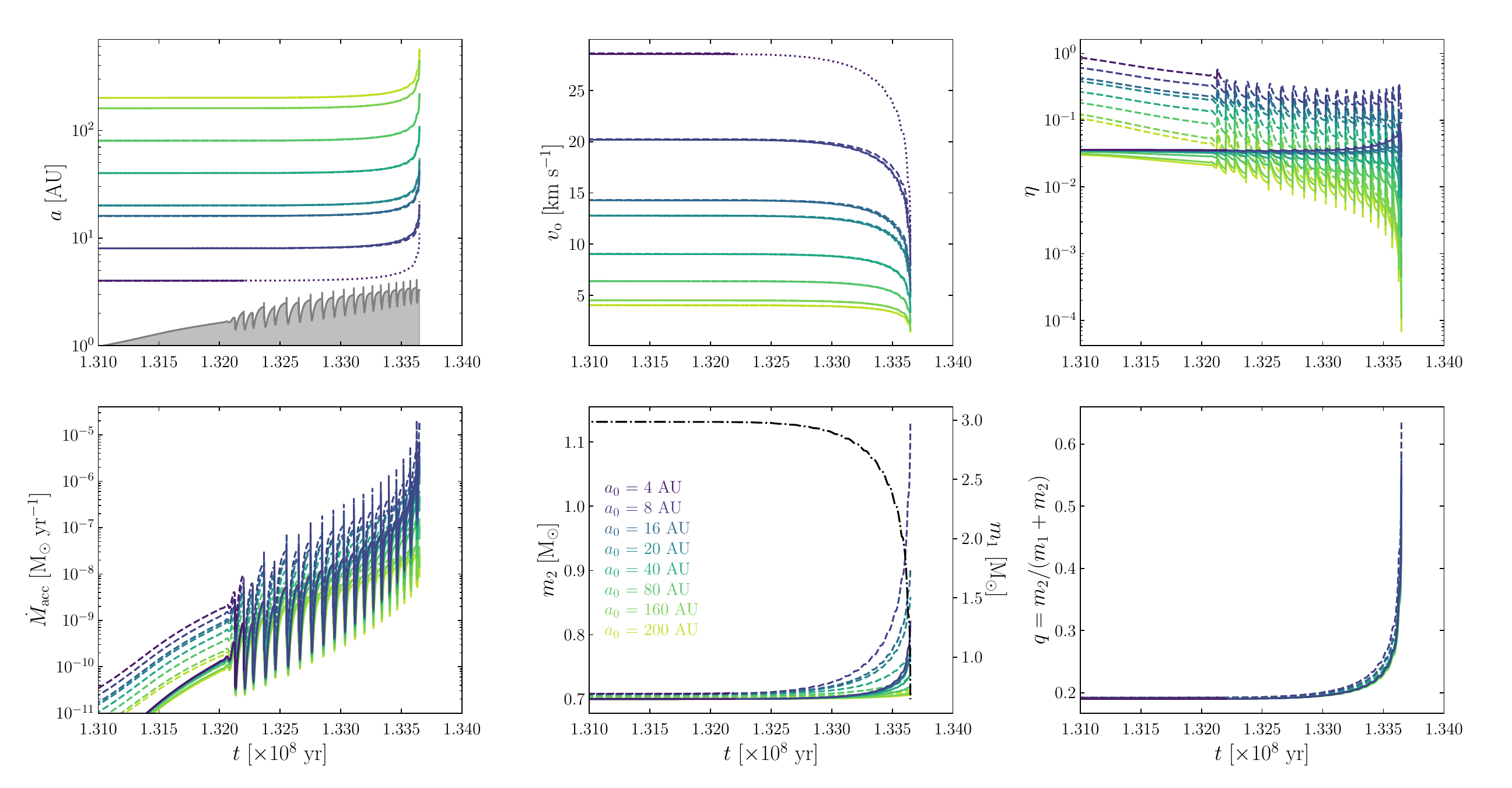}
\end{center}
\caption{The same as Fig.~\ref{fig:time_evol_1M_zoom} but for models with initial configuration of $m_{1,0}=3$ M$_\odot$ and $m_{2,0}=0.7$ M$_\odot$.}
\label{fig:time_evol_3M_zoom}
\end{figure*}

Fig.~\ref{fig:time_evol_1M} provides a representative example of the time evolution of a binary with initial masses $m_{1,0} = 1$ M$_\odot$, $m_{2,0} = 0.7$ M$_\odot$, showcasing the results for all eight initial separations and the three types of accretion prescriptions. This figure illustrates the time evolution of key parameters: binary separation ($a$), orbital velocity ($v_\mathrm{o}$), mass-accretion efficiency ($\eta$), mass accretion rate onto the WD ($\dot{M}_\mathrm{acc}$), WD mass ($m_2$), and the dimensionless mass ratio ($q$). The time origin ($t=0$) corresponds to the onset of the RGB phase $t_\mathrm{iRGB}$ (see Table~\ref{tab:models}).

\begin{figure}
\begin{center}
\includegraphics[width=\linewidth]{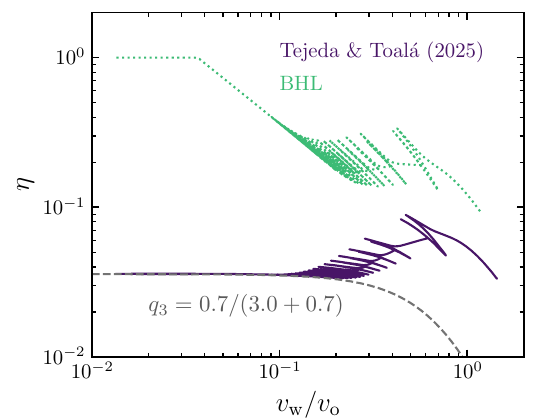}
\end{center}
\caption{Evolution of the mass accretion efficiency $\eta$ versus $w (= v_\mathrm{w}/v_\mathrm{o})$ for an accreting WD with $m_{2,0}=0.7$ M$_\odot$ orbiting an evolving $m_{1,0}=3.0$ M$_\odot$ at an initial orbital separation of $a_0=8$ AU. The solid and dotted lines show the results adopting the accretion from Paper I and that of the standard BHL accretion, respectively. The dashed curve ($q_3$) represent the analytical calculations for $\eta$ adopting the initial conditions of the model.}
\label{fig:eta_space_single}
\end{figure}

\begin{figure*}
\begin{center}
\includegraphics[width=0.8\linewidth]{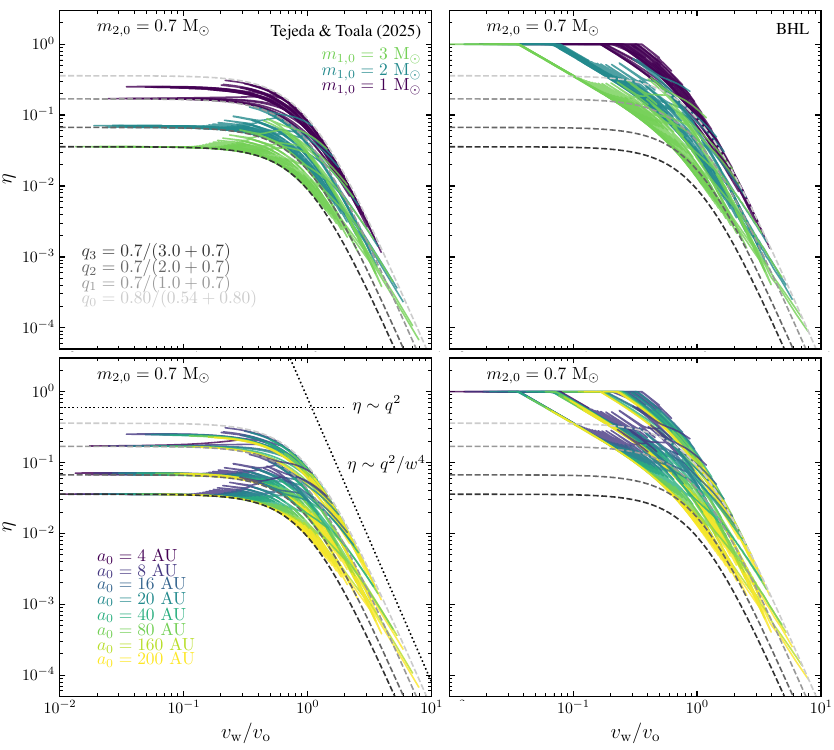}
\end{center}
\caption{Evolution of the mass accretion efficiency $\eta$ versus $w (= v_\mathrm{w}/v_\mathrm{o})$ diagram. The left column panels show the results from models with accretion model from Paper I, while the right column panels show the results of applying the standard BHL formalism. The dashed curves represent the analytical calculations for $\eta$ adopting the initial conditions of the models ($q_1$, $q_2$, and $q_3$) and the final condition of the $m_{1,0}=1$ M$_\odot$ models ($q_0$). The top panels illustrate the evolution of the models highlighting the initial mass of the primary star $m_{1,0}$ while the bottom panels highlight the evolution of the models with different initial orbital separations $a_0$. The dotted lines in the bottom-left panel show the two asymptotic behaviours of Eq.~(\ref{eq:TejedaToala}).}
\label{fig:eta_space2}
\end{figure*}

Simulations without accretion serve as the baseline for comparison. Due to continuous mass loss from the primary star, these systems experience orbital expansion. This is evident in the upper-left and upper-middle panels of Fig.~\ref{fig:time_evol_1M}, which show an increasing orbital separation ($a$) and a decreasing orbital velocity ($v_\mathrm{o}$).

In contrast to the no-accretion scenario, from this figure we can also appreciate the significant role accretion plays in shaping the evolution of symbiotic systems, particularly during the transition from the RGB to the TPAGB phases. The initial orbital separation also emerges as a key factor influencing the final outcomes and accretion efficiencies. Notably, models incorporating accretion (solid and dashed lines) exhibit shorter final orbital separations than those without accretion, an effect amplified by smaller initial orbital separations and more prominent during the TPAGB phase.

Furthermore, the choice of accretion model impacts the system's evolution. The standard BHL formalism (dashed lines) predicts larger mass accretion efficiencies ($\eta$) than the modified model from Paper I (solid lines), as shown in the top-right panel of Fig.~\ref{fig:time_evol_1M}. Consequently, the orbits of systems modelled with the standard BHL model are less, as the total mass of the system is better conserved due to the higher $\eta$. Fig.~\ref{fig:time_evol_1M_zoom} provides a closer look at the final TPAGB phase for the \mbox{$m_{1,0}=1$ M$_\odot$} model, highlighting these differences. Note that the case with the most compact configuration does not appear in this figure, because the model reached the condition for WRLO and was stopped at the expansion of the RGB phase (see Fig.~\ref{fig:time_evol_1M}).

Models with more massive donor stars exhibit similar evolutionary trends, although less pronounced, as shown in Fig.~\ref{fig:time_evol_2M_zoom} and \ref{fig:time_evol_3M_zoom} for $m_{1,0} = 2$ and 3 M$_\odot$, respectively. Note that also in those models, the most compact simulation is stopped at the begging of the TPAGB phase as they reached the condition for a WRLO evolution (see Appendix~\ref{app:WRLO}). Simulations with a WD of initial mass $m_{2,0}=1$ M$_\odot$ exhibit comparable evolutionary trends as those presented in Figs.~\ref{fig:time_evol_1M}--\ref{fig:time_evol_3M_zoom}.

Across all models, the simulations including the modified wind accretion from Paper I predict mass accretion rates $\dot{M}_\mathrm{acc}$ between $10^{-10}$ and $\lesssim 10^{-5}$~M$_\odot$~yr$^{-1}$ during the TPAGB phase (relevant for S-type symbiotic systems), as shown in the bottom-left panels of Figs.~\ref{fig:time_evol_1M_zoom}--\ref{fig:time_evol_3M_zoom}. These accretion rates evidently reflect the pulsating behaviour of the corresponding evolving stars, with $\dot{M}_\mathrm{acc}$ increasing in tandem with the stellar mass-loss rate $\dot{M}_\mathrm{w}$ (see Fig.~\ref{fig:vel_wind}). Notably, the more massive models exhibit higher mass-loss rates (see Fig.~\ref{fig:mdot}), but their dynamical response is tempered by lower mass accretion efficiencies $\eta$ (top-right panels of Figs.~\ref{fig:time_evol_1M_zoom}--\ref{fig:time_evol_3M_zoom}).

Depending on the stellar evolution model, the TPAGB phase is not the only stage characterised by significant mass-loss rate. In particular, the 1 M$_\odot$ stellar evolution model exhibits substantial mass ejection during the RGB phase (D-type systems), just before the He-flash (see Fig.~\ref{fig:mdot}). The bottom-left panel of Fig.~\ref{fig:time_evol_1M} reveals that during the RGB phase, $\dot{M}_\mathrm{acc}$ is comparable to values predicted for the TPAGB phase. Particularly, the $m_{1,0}=1$ M$_\odot$ stellar evolution model produces the ejection of about 0.3 M$_\odot$ during the RGB phase (Fig.~\ref{fig:time_evol_1M}, bottom-middle panel). Such behaviour, however, is less pronounced in the more massive stellar evolution models.

The mass accretion efficiency history shown in Figs.~\ref{fig:time_evol_1M}--\ref{fig:time_evol_3M_zoom} illustrates that the highest accretion rates are attained in systems with the smallest initial orbital separations.
This is a natural consequence of the dependence of $\eta$ on the velocity ratio $w = v_\mathrm{w}/v_\mathrm{o}$ (see Eq.~\ref{eq:TejedaToala} and \ref{eq:BHL}), where closer orbital separations result in higher orbital velocities and thus smaller values of $w$. In addition, models with larger $m_{1,0}$ have lower accretion efficiencies due to their smaller mass ratios $q$.

Fig.~\ref{fig:eta_space_single} shows the evolution of a representative system with initial masses $m_{2,0}=0.7$ M$_\odot$ and $m_{1,0}=3.0$ M$_\odot$, and an initial orbital separation of $a_0=8$ AU in the $\eta$ vs.~$w$ parameter space. The evolutionary tracks are illustrated for both the modified accretion prescription from Paper I and the standard BHL model. 
The main trend for both prescriptions is an increasing wind-to-orbital velocity ratio $w$ (evolving from left to right in the figure). However, this trend can temporarily reverse during the TPAGB phase in response to the thermal pulses of the primary star, as the system undergoes episodes of increased mass loss and enhanced wind velocities. 
The BHL model consistently predicts higher accretion efficiency than the Paper I model, with the difference exceeding an order of magnitude at the start of the evolution. While the BHL model generally shows a decreasing accretion efficiency, the Paper I model exhibits a more complex behaviour. It starts with a constant value $\eta \simeq q^2<1$, then increases during the TPAGB phase, eventually reaching a turnover point and declining slightly.

Fig.~\ref{fig:eta_space2} summarizes the evolution of all accreting models in the  $\eta$ vs.~$w$ parameter space. The left panels display the results for the Paper I model, and the right panels show those for the standard BHL model. In the top row, colours distinguish models with different initial primary masses, while in the bottom row, colours differentiate models with different initial orbital separations. As a reference, dashed lines in all panels show the $\eta-w$ evolution for fixed $q$ corresponding to the initial configuration of the simulations ($q_1$, $q_2$, and $q_3$). The parameter $q_0$ corresponds to the final configuration of the \mbox{$m_{1,0}=1$\,M$_\odot$} simulations when the accretor reaches 0.8\,$\mathrm{M_\odot}$.

The evolutionary tracks in the left panels of Fig.~\ref{fig:eta_space2} generally resemble those described in Fig.~\ref{fig:eta_space_single} for initial distances $a_0\leq16\,$AU. However, models with larger $a_0$ do not exhibit a marked turnover, but rather follow a decreasing mass accretion efficiency trend, asymptotically approaching $\eta \approx q^2 / w^4$ (Eq. \ref{eq:TejedaToala}) in the large $w$ limit, where the Paper I model approximates the standard BHL theory. The early evolution of all models occurs in the region where $\eta \approx q^2$ for small $w$, dominated by the modified wind accretion formulation. These two limiting behaviours are illustrated in the bottom-left panel of Fig.~\ref{fig:eta_space2} with dotted lines.

The right-column panels of Fig.~\ref{fig:eta_space2} show the results for the standard BHL accretion approximation. As demonstrated in Paper I, the calculations for $w > 1$ are consistent between the two accretion prescriptions, but for $w < 1$, the BHL calculations yield non-physical $\eta_\mathrm{BHL}$ values larger than 1. To avoid these non-physical values, which have been extensively discussed in previous works (see references in Section \ref{sec:intro}), these cases were restricted to $\eta_\mathrm{BHL}=1$. Notably, all models using the standard BHL accretion approximation exhibit larger $\eta$ values than those with the same configuration but adopting the accretion regime from Paper I, even in the $w > 1$ regime.

\section{Discussion}
\label{sec:discussion}

By incorporating the accretion formalism proposed in Paper I, we were able to investigate the evolution of symbiotic systems and their accretion history throughout the lifetimes of both stars in the system, in realistic accretion efficiency regimes.
Our results indicate a general trend: symbiotic systems with small initial separations $a_0$ or systems early in their evolution exhibit the $\eta \sim q^2$ asymptotic behaviour. Over time, and best seen in systems with large $a_0$, the systems transition to the $\eta \sim q^2 / w^4$ regime, characteristic of the standard BHL formulation for wind accretion in binary systems for large $w$ values.

We remark that models with $w (= v_\mathrm{w}/v_\mathrm{o}) < 1$ do not necessarily enter the WRLO. According to \citet{Mohamed2012}, if the wind launching radius is smaller than the Roche Lobe, the accretion is purely due to wind capture. This situation is reached by the evolving systems with $a_0 = 4$ AU and, thus, their further evolution is not followed. Consequently, the numerical results presented here correspond to purely to the wind accretion regime.

\begin{figure}
\begin{center}
\includegraphics[width=0.95\linewidth]{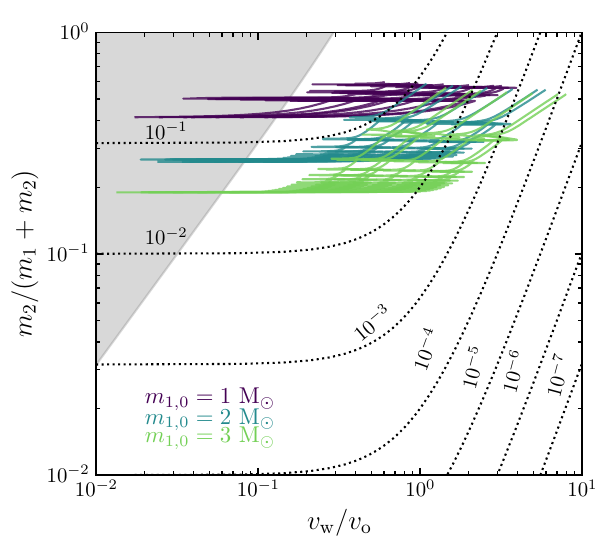}\\
\includegraphics[width=0.95\linewidth]{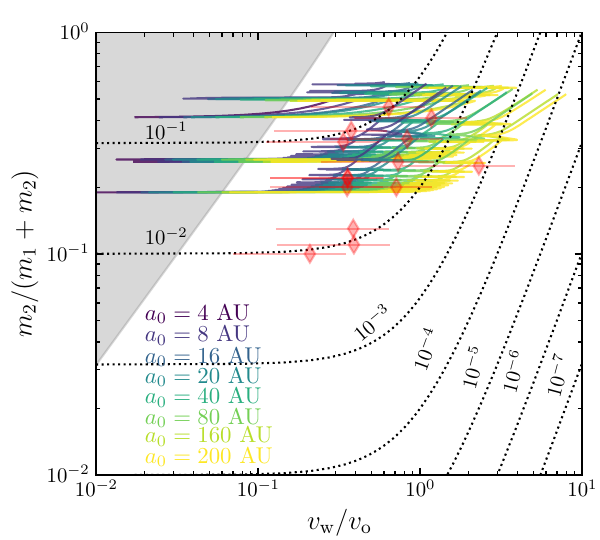}
\end{center}
\caption{Evolution of the symbiotic systems with $m_{1,0}=1$, 2 and 3 M$_\odot$ and an accreting companion with $m_{2,0}=0.7$ M$_\odot$ (colours) in the $q=m_2/(m_1+m_2)$ versus $w=v_\mathrm{w}/v_\mathrm{o}$ parameter space. Dotted lines indicate contours of constant $\eta$ values. The diamonds in the bottom panel shows the expected position of symbiotic systems listed in Appendix~\ref{app:symbiotics}.}
\label{fig:eta_space}
\end{figure}

A WD with initial mass of $m_{2,0} = 0.7$ M$_\odot$ orbiting mass-losing stars with initial masses of $m_{1,0} = 1$, 2, and 3 M$_\odot$ result in mass accretion rates ranging from $\dot{M}_\mathrm{acc} = [10^{-10}$--$10^{-5}$]~M$_\odot$~yr$^{-1}$. These rates increase significantly toward the end of the TPAGB phase where the mass-loss rate of the donor star is the highest. While this range is broad, the predicted $\dot{M}_\mathrm{acc}$ values align well with estimates derived from optical and X-ray observations of symbiotic systems \citep[e.g.,][]{Pujol2023,VT2024,Zhekov2019,Guerrero2025,Marchev2024,Boneva2021,Luna2018}.
However, determining the mass accretion efficiency $\eta$ is more challenging. Accurate measurements require precise estimates of $\dot{M}_\mathrm{w}$, which depend on detailed characterization of the late-type companion in the system and/or dedicated observational campaigns \citep[e.g.,][]{Ramstedt2018}.

The simulations predict that symbiotic systems  accreting through a wind capture mechanism have mass accretion efficiencies in the $\eta= [10^{-4} - 10^{-1}]$ range as illustrated in the $q$--$w$ space of Fig.~\ref{fig:eta_space}, where we also plot contours of $\eta$.
The bottom panel in this figure display the expected locations of a sample of symbiotic systems listed in Appendix~\ref{app:symbiotics}, with data extracted from the {\it New Online Database of Symbiotic Variables} \citep{Merc2019}. For these systems, a standard wind velocity of $v_\mathrm{w} = 12\pm 8$~km~s$^{-1}$ is adopted, consistent with reported wind velocities for late-type stars \citep[see][]{Ramstedt2020}. Fig.~\ref{fig:eta_space} shows a general agreement of the observed properties of symbiotic systems and the models presented here including the modified accretion scheme of Paper I. Notably, the system $o$ Ceti stands out, having one of the most extended orbits with a period of 498 yr. This results in a small orbital velocity $v_\mathrm{o}$ and, consequently, a large $w$, placing it as the rightmost data point in the figure.

The panels in Fig.~\ref{fig:eta_space} include a gray-shaded region representing the part of the $q$--$w$ space where the analytical approximation of Paper I is not applicable. That work assumes that the mass-losing star $m_1$ returns to its unperturbed state between consecutive passages of the accretor $m_2$. To evaluate this condition, Paper I introduced a refill time ($t_\mathrm{fill}$), which must be smaller than the system's orbital period $T$. For circular orbits, this condition can be expressed as:
\begin{equation}
    \frac{t_\mathrm{fill}}{T} = \frac{q}{\pi} (w^3+w)  < 1.
\end{equation}

Fig.~\ref{fig:eta_space} shows that most models evolve while avoiding the non-valid region of the $q$--$w$ parameter space.  
All models evolve from within the non-valid region of the $q$--$w$ diagram due to the extremely low stellar wind velocities ($v_\mathrm{w} < 2$~km~s$^{-1}$) adopted prior to the onset of the peak of the RGB phase (see Fig.~\ref{fig:vel_wind}). A notable exception is seen in the models with $m_{1,0} = 1$ M$_\odot$, which re-enter the non-valid region of the $q$--$w$ diagram during the evolutionary stage between the He-flash and the onset of the TPAGB phase. This occurs because the wind velocity decreases significantly during this interval (Fig.~\ref{fig:vel_wind}, top left panel). We predict that during these specific evolutionary phase, no significant mass is accreted, because $\dot{M}_\mathrm{w}$ is extremely low ($< 10^{-11}$ M$_\odot$ yr$^{-1}$) making the impact of this intermediate phase negligible.

\begin{table}
\begin{center}
\caption{Final masses at the end of the simulations. $m_\mathrm{F,TT}$ and $m_\mathrm{F,BHL}$ are the final masses for the modified wind accretion from Paper I and the standard BHL accretion, respectively. Values denoted with $^\star$ represent models that entered the WRLO.}
\footnotesize
\setlength{\tabcolsep}{\tabcolsep}  
%\scriptsize
\begin{tabular}{lcccc}
\hline
$m_{1,0}$    & $m_{2,0}$   &  $a_0$  &  $m_\mathrm{F,TT}$ & $m_\mathrm{F,BHL}$ \\
(M$_\odot$)  & (M$_\odot$) &  (AU)   & (M$_\odot$)        &  (M$_\odot$) \\
\hline 
1   &   0.7   &  4    &  0.748$^\star$  &  0.873$^\star$\\
    &         &  8    &  0.785  &  1.067\\
    &         &  16   &  0.768  &  0.934\\
    &         &  20   &  0.762  &  0.900\\
    &         &  40   &  0.744  &  0.821\\
    &         &  80   &  0.728  &  0.773\\
    &         &  160  &  0.718  &  0.744\\
    &         &  200  &  0.715  &  0.738\\
    & 1.0     &  4    &  1.050$^\star$  &  1.151$^\star$\\ 
    &         &  8    &  1.125  &  1.400\\
    &         &  16   &  1.104  &  1.330\\
    &         &  20   &  1.096  &  1.287\\
    &         &  40   &  1.070  &  1.179\\
    &         &  80   &  1.046  &  1.108\\
    &         &  160  &  1.029  &  1.065\\
    &         &  200  &  1.025  &  1.056\\
\hline
2   &   0.7   &  4  &    0.705$^\star$  &  0.762$^\star$\\
    &   	  &  8  &    0.819  &  1.161\\    
    &	      & 16  &    0.779  &  0.924\\
    &	      & 20  &    0.768  &  0.878\\
    &	      & 40  &    0.739  &  0.788\\
    &  	      & 80  &    0.720  &  0.745\\
    &	      & 160 &    0.710  &  0.723\\
    &	      & 200 &    0.708  &  0.718\\

    &   1.0   & 4   &   1.008$^\star$   &   1.066$^\star$\\
    &   	  & 8   &   1.200 &   1.400\\    
    &	      & 16  &   1.140 &   1.400\\
    &	      & 20  &   1.121 &   1.327\\
    &	      & 40  &   1.072 &   1.160\\
    &	      & 80  &   1.038 &   1.079\\
    &	      & 160 &   1.019 &   1.040\\
    &	      & 200 &   1.015 &   1.033\\
\hline
3  &    0.7  & 4    &  0.700$^\star$  &  0.709$^\star$\\
   &    	  & 8    &  0.820  &  1.133\\   
   &	      & 16   &  0.779  &  0.902\\
   &	      & 20   &  0.767  &  0.859\\
   &	      & 40   &  0.738  &  0.776\\
   &	      & 80   &  0.720  &  0.734\\
   &	      & 160  &  0.709  &  0.715\\
   &	      & 200  &  0.707  &  0.711\\
   &  1.0    & 4    &  1.001$^\star$  &  1.010$^\star$ \\
   &     	  & 8    &  1.217  &  1.400 \\   
   &	      & 16   &  1.147  &  1.400 \\
   &	      & 20   &  1.126  &  1.313 \\
   &	      & 40   &  1.074  &  1.144 \\ 
   &	      & 80   &  1.038  &  1.066 \\
   &	      & 160  &  1.017  &  1.029 \\
   &	      & 200  &  1.013  &  1.022 \\
\hline
\end{tabular}
\label{tab:times}
\end{center}
\end{table}

\subsection{Reaching the Chandrasekhar limit}
\label{sec:Ch_limit}

One of the major differences between the two wind accretion schemes used in the simulations is the final mass of the accretor. None of the simulations adopting the modified wind accretion model from Paper I reached the Chandrasekhar limit, not even those with $m_{2,0}=1.0$ M$_\odot$. Table~\ref{tab:times} lists the final masses for all model configurations.

Simulations with smaller initial orbital separations result in a higher final WD mass with the exception of binaries with $a_0=$ 4, for which the simulations stopped due to reaching when reaching the WRLO condition ($R_\mathrm{RL}=R_\mathrm{cond}$; see Appendix~\ref{app:WRLO}) and were stopped. In the case of simulations that include the standard BHL accretion scheme, the 1.4 M$_\odot$ mass limit is achieved for cases with initial mass of the accretor of $m_{2,0}=1.0$ M$_\odot$ for initial orbital configurations smaller than 20 AU. On the other hand, larger initial orbital separations predict very similar final masses for the accretor in both accretion regimes, reflecting the fact that those simulations fall within the $w >1$ regime, which is where both schemes approximate to similar results.

The significant differences between the two wind accretion prescriptions could substantially impact binary and population synthesis simulations. While incorporating efficiency factors into the standard BHL model might improve accuracy for compact systems ($w < 1$), it can lead to underestimation of the accretor's final mass in wider systems ($w > 1$). New binary and population synthesis calculations should prioritize incorporating more accurate wind accretion models, such as the one proposed in Paper I.

The final masses of the accreting WDs at the end of the simulations, summarized in Table \ref{tab:times}, suggest that Type Ia supernovae can only occur in symbiotic systems with the modified wind accretion model from Paper I if the initial mass of the accreting WD is relatively large, exceeding 1.0 M$_\odot$. Such WDs are classified as ultra-massive and are typically formed from the evolution of Solar-like stars with initial masses greater than 7 M$_\odot$ \citep{Camisassa2019,Curd2017,Hermes2013,Vennes2008}. This implies that Type Ia supernovae arising from symbiotic systems would be relatively rare. However, ultra-massive WDs can also be formed through the merger of double degenerate systems, which would lead to more complex evolutionary pathways \citep{GarciaBerro1997,GarciaBerro2012}.
\\

\begin{figure*}
\begin{center}
\includegraphics[width=\linewidth]{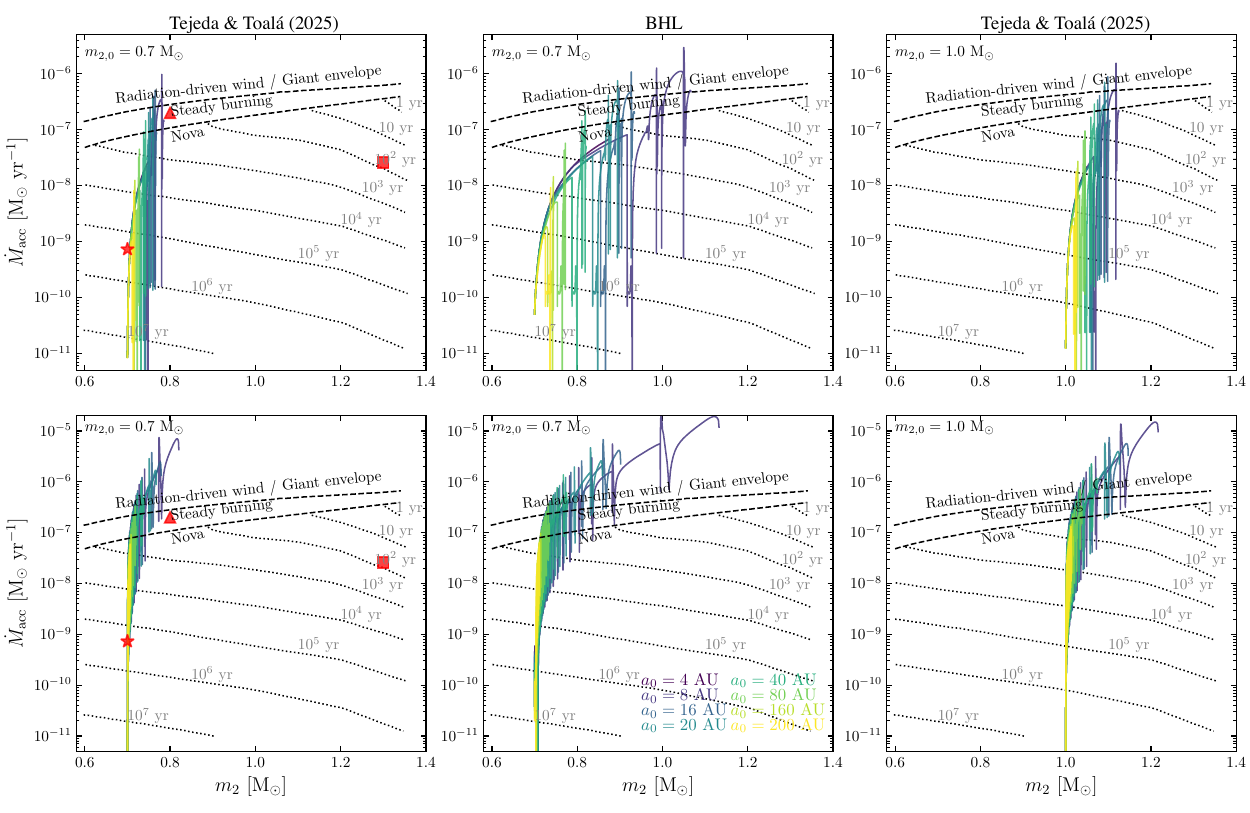}
\end{center}
\caption{Mass accretion rate $\dot{M}_\mathrm{acc}$ as a function of the mass of the accreting WD $m_2$. The top and bottom panels show the results from simulations with $m_{1,0} = 1$ and 3~M$_\odot$, respectively. The left- and right-column panels show results adopting the accretion from Paper I, while the middle column panels are those with the standard BHL accretion. This figure was adapted from \citet{Chomiuk2021} which adapted it from results presented in \citet{Wolf2013}. Dotted lines show constant nova recurrence times. Dashed lines show the limits between the three accretion regimes. Different coloured lines show the evolution of simulations with different initial orbital separation. The star, triangle, and square symbols in the left panels represent the positions of the symbiotic systems R Aqr, Y Gem, and T CrB, respectively (see Section~\ref{sec:evolution_accretor} for details).}
\label{fig:accretion_rate}
\end{figure*}

\subsection{The evolution of the accretor}
\label{sec:evolution_accretor}

In general, during the evolution of accreting WDs H-rich material accumulates on the surface. This material undergoes nuclear burning, eventually becoming part of the WD. There are three distinct accretion regimes \citep[see][and references therein]{Nomoto2007,Shen2007,Wolf2013,Chomiuk2021}: {\it i}) Steady Burning: This occurs when the rate of accreted H-rich material is balanced by nuclear burning on the WD's surface, allowing for a stable accumulation of material. {\it ii}) High Accretion Rate: At higher accretion rates, material accumulates on the WD’s surface faster than it can burn. This may lead to the formation of an extended envelope resembling a red giant, or the material could be ejected through a radiation-driven wind \citep[e.g.,][]{Hachisu1996}. {\it iii}) Low Accretion Rate: At lower accretion rates, nuclear burning becomes unstable, resulting in a thermonuclear runaway event that produces a nova. Nova events are produced by accreting WDs in cataclysmic variable systems, where accretion takes place in a Roche lobe overflow interaction with a RSG \citep{Starrfield1987}, but they can also occur in symbiotic systems \citep{Kenyon1983} where accretion is produced via the companion's wind (as modelled in the present work) or a WRLO scenario.

Fig.~\ref{fig:accretion_rate} illustrates the evolution of the mass accretion rate as a function of the evolving mass of the accretor for simulations with $m_{1,0} = 1$ (top) and 3 M$_\odot$ (bottom), for accretion regimes determined in Paper I and the standard BHL wind accretion. The panels also show regions corresponding to different nova recurrence times (dotted lines) and the boundaries between the three accretion regimes (dashed lines). We found that due to the highly variable nature of the donor star's mass-loss rate during the TPAGB phase, all simulations make the evolving symbiotic systems to transition back and forth between different nova recurrence times. Particularly, models with smaller initial orbital configurations make the accretor to transition between the three accretion regimes. 
The left- and right-column panels of Fig.~\ref{fig:accretion_rate}, showing models with accretion from Paper I, corroborate that under such assumption none of the accreting WDs reach the Chandrasekhar limit. In those cases, the net accreted mass is about 0.1--0.2 M$_\odot$ depending on the model (see Table~\ref{tab:times}). 
In general, models including the standard BHL accretion reach larger final WD masses. In particular, models with $m_{2,0}=1$ M$_\odot$ reach the Chandrasekhar mass limit for some $a_0$ (see Table~\ref{tab:times}).

We remark that the models spend most of their evolution between the nova and steady burning phases. It is only when experiencing the last thermal pulses that the accretor remains in the giant envelope accretion regime, when the mass-loss rate from the donor star is highest. This situation is best seen for the $m_{1,0}= 3$ M$_\odot$ simulations given that the mass-lost rate of the donor star can reach values as high as 10$^{-4}$ M$_\odot$ yr$^{-1}$ in the last thermal pulses (Fig.~\ref{fig:mdot}, bottom right panel). The time spend in this last evolving sequence is short and we expect a very small number of systems experiencing the giant envelope accreting regime, before the mass-losing star evolves into the post-AGB phase to finally become a WD star itself. Ultimately, producing a double-degenerate system.

For comparison, the panels in the left column of Fig.~\ref{fig:accretion_rate} include the estimated parameters for three different cases of accreting symbiotic systems. The S-type system R Aqr, has an extended orbit \citep[$a = 14.5$ AU;][]{Alcolea2023} and an estimated mass accretion rate of $7.3 \times 10^{-10}$~M$_\odot$~yr$^{-1}$ \citep{VT2024}. According to the models presented here, this system has an approximate recurrence time for nova eruptions of $\sim10^{6}$ yr. Recently, \citet{Guerrero2025} identified the X-ray-emitting AGB star Y Gem as a new S-type symbiotic system with one of the highest UV fluxes observed. They demonstrated that this system is currently in a steady burning phase. Our models are also consistent with its position in the diagrams of Fig.~\ref{fig:accretion_rate}. Finally, we also show the position of the recurrent symbiotic nova system T CrB, which has a massive accreting WD \citep[$\approx 1.3$~M$_\odot$;][]{Stanishev2004,Shara2018}, placing it close to the Chandrasekhar limit. The mass accretion rate recently estimated by \citet{Toala2024} positions this system near a recurrence time of $\sim 10^2$ yr, consistent with the 80-yr recurrence period recorded for this nova system \citep{Schaefer2023}. We note that this S-type symbiotic system has a relatively short orbital period of about 230 d, which strongly suggest that its accretion mechanism is through a WRLO scenario. However, a comparison with this systems is of relevance, because the simulations presented in this work suggest that none of the wind accreting WDs will reach such high masses unless the accreting component was already an ultra-massive WD prior to accreting material from its evolving companion. 

\begin{figure*}
\begin{center}
\includegraphics[width=0.4\linewidth]{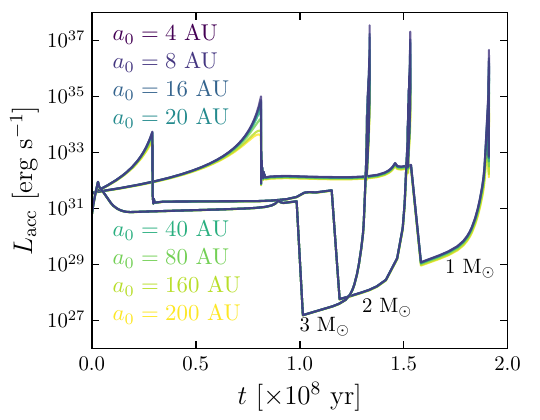}~
\includegraphics[width=0.4\linewidth]{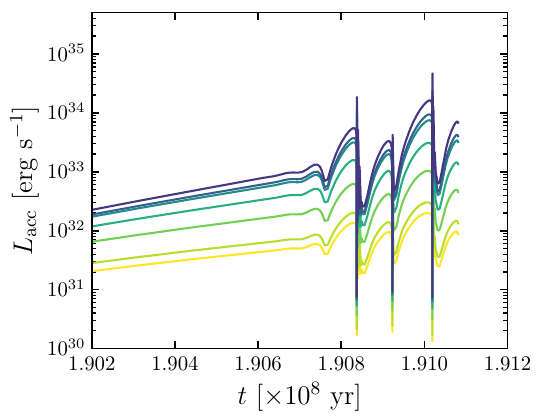}\\
\includegraphics[width=0.4\linewidth]{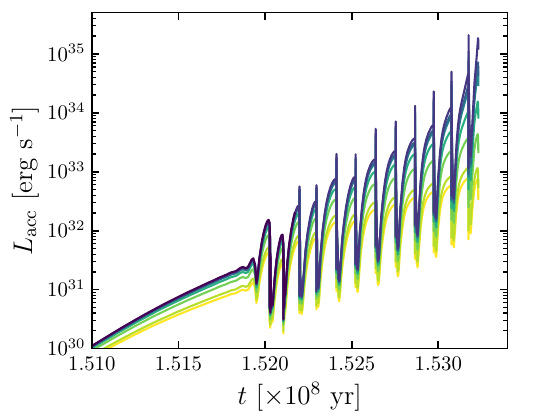}~
\includegraphics[width=0.4\linewidth]{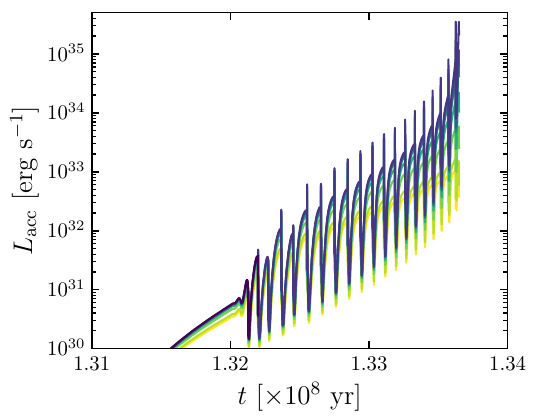}
\end{center}
\caption{Evolution of the bolometric luminosity produced by accretion $L_\mathrm{acc}$ with time $t$ for models with accretion from Paper I. The top left panel shows all modes, whilst other panels show the TPAGB phase of the different models.}
\label{fig:Lum}
\end{figure*}

It is important to remark that the models presented here do not account for mass ejection during nova events. However, we suggest that this omission is unlikely to significantly impact our calculations, as the mass lost per nova event is expected to be less than $10^{-4}$~M$_\odot$ \citep{Bode2010} and decreases as the mass of the accretor increases \citep[see figure 14 in][]{Wolf2013}. Self-consistent models of nova eruptions have been presented in the literature, but mostly for massive WDs \citep[e.g.,][]{Kato2017,Kato2022,Kato2024}. Other studies have calculated mass accretion in evolving symbiotic systems while modelling nova events and their associated mass ejection under the wind accretion scheme. For instance, \citet{HillmanKashi2021} and \citet{Vathachira2024} modelled systems with WDs of initial masses 0.7, 1.0, and 1.25 M$_\odot$ accreting material from a companion with a mass of 1 M$_\odot$ evolving through the late TPAGB phase. Using the same stellar evolution models and codes, both works suggest that massive WDs or accretors located at orbital separations greater than 20 AU are likely to lose more mass through nova ejections than they accrete, implying that such systems are unlikely to evolve into type Ia supernova progenitors. Conversely, \citet{Vathachira2024} demonstrated that systems hosting less massive WDs or those at closer orbital separations exhibit net mass accretion rates that exceed the total mass ejected during nova events. Notably, these studies applied a standard BHL wind accretion formalism in the $w >1$ regime and reported mass accretion efficiencies $\eta = \dot{M}_\mathrm{acc}/\dot{M}_\mathrm{w} < 0.02$, which is an order of magnitude lower than our predictions for a mass-losing star with $m_{1,0} = 1$~M$_\odot$ in similar orbital configuration as those authors (Fig.~\ref{fig:time_evol_1M_zoom}, top right panel)\footnote{Although not clear in their publication, \citet{Vathachira2024} seem to adopt a constant wind velocity for their late-type star of 20 km~s$^{-1}$.}.
Future work will extend our calculations by incorporating the modified wind accretion formalism proposed in Paper I alongside the effects of mass ejection during nova events. Those efforts will be addressed in a subsequent paper.

\subsection{Bolometric Luminosity}

We used the results from the simulations to make predictions on the bolometric luminosity produced by the accretion process. We converted the accreted mass into radiation adopting a thin disc model. The bolometric luminosity produced by the accretion process can be expressed as \citep[e.g.,][]{Shakura1973,Pringle1981}
\begin{equation}
    L_\mathrm{acc} = \frac{1}{2}\frac{G m_2}{R_2} \dot{M}_\mathrm{acc},
\end{equation}
\noindent where $R_2$ is the radius of the accreting WD.

The evolution with time of the bolometric luminosity $L_\mathrm{acc}$ is presented in Fig.~\ref{fig:Lum}, where we have adopted a conservative radius for the WD of $R_2 = 0.01$ R$_\odot$, consistent for a WD mass of 0.7 M$_\odot$ \citep[e.g.,][]{Boshkayev2016,Pasquini2023,Karinkuzhi2024}. The top left panel of Fig.~\ref{fig:Lum} shows the complete evolution of $L_\mathrm{acc}$ following the stellar evolution models including the wind accretion from Paper I, while other panels show the corresponding TPAGB phases of the different evolving models.

$L_\mathrm{acc}$ represents the total budget for radiation produced by the accretion process that should contribute to all wavelengths. Detailed studies of accreting symbiotic systems, such as those presented for R Aqr and Y Gem, showed that the X-ray emission is only a fraction of the observed optical emission produced by the disc. \citet{Guerrero2025} and \citet{VT2024} found that for these systems $L_\mathrm{X} \approx$ [0.004--0.07]$L_\mathrm{acc}$. Taking this efficiency factor, the estimated $L_\mathrm{X}$ from the $L_\mathrm{acc}$ values in Fig.~\ref{fig:Lum} agree well with the large number of X-ray observations of accreting symbiotic systems in the literature \citep[e.g.,][]{Luna2013,Lima2024,Guerrero2024_XAGBs} with estimated $L_\mathrm{X}$=[10$^{30}$--10$^{34}$]~erg~s$^{-1}$, the rest of the emission seem to be dominated by the optical emission of the accretion disc. Note that even during the RGB phase of the $m_{1,0} = 1.0$ M$_\odot$ simulations, the predicted luminosity is also comparable to that estimated from observations.

\section{Summary}
\label{sec:summary}

For decades, numerical simulations found that the standard implementation of the BHL wind accretion model for binary systems did not produce reliable results for the case where the stellar wind velocity $v_\mathrm{w}$ was smaller than the orbital velocity of the accretor $v_\mathrm{o}$. A situation that is of outmost importance for symbiotic systems. The standard BHL scheme predicts non-physical mass accretion efficiencies larger than 1 for $v_\mathrm{w} < v_\mathrm{o}$. Recently, \citet{TejedaToala2025} (Paper I)  proposed a geometric correction to the standard implementation of the BHL model for accreting binaries and found a better agreement between their analytical predictions and numerical simulations.

In this paper, we studied the impact of the wind accretion model of Paper I  for the case of evolving symbiotic systems. We modelled 48 different symbiotic systems, each consisting of a WD with an initial mass of  $m_{2,0} = 0.7$ or 1.0 M$_\odot$ and a Solar-like star with an initial mass of $m_{1,0}=1$, 2, and 3 M$_\odot$. The stars were placed in initially circular orbits with separations ranging from 4 to 200 AU. The evolution of the Solar-like stars was computed using the stellar evolution code MESA, and the orbital evolution of the systems was calculated making use of the N-body package REBOUND. For each configuration, we conducted three types of simulations: accretion using the modified model from Paper I, accretion using the standard BHL model, and no accretion. The simulations are restricted to systems accreting purely through wind capture, and systems entering the WRLO phase are not discussed in the present work.

Our main findings are summarized as follows:
\begin{itemize}

    \item The wide range of masses and orbital configurations modelled highlights distinct evolutionary paths. The adopted accretion prescription significantly affects the final destiny of the modelled symbiotic system. None of the simulations including the wind accretion from Paper I reached the Chandrasekhar limit (1.4 M$_\odot$). For this accretion model to produce recurrent nova and type Ia supernova events in symbiotic systems, the accreting WD should start with an initial mass in the range known for ultra-massive WDs. In other words, that the WD should be the descendent of a massive ($> 7$ M$_\odot$) progenitor. On the other hand, compact systems ($a_0 < 20$ AU) modelled with the standard BHL scheme evolve ultimately taking the accreting WDs to this mass limit for the two adopted masses of the accretor. This discrepancy suggests that binary simulations and population synthesis studies relying on the BHL approximation may require revision, even in cases where arbitrary efficiency factors are adopted.

    \item Simulations including the wind accretion from Paper I generally agree with binary and orbital properties of a number of observed symbiotic systems. The accretion properties of well-characterised S-type symbiotic binaries with wide orbits, such as R Aqr and Y Gem, are nicely reproduced by the wind accretion scheme of Paper I. The nova recurrence time of R Aqr is estimated to be a few times 10$^{6}$~yr.

    \item Simulations predict that accretors can transition back and forth through different recurrence nova times, and even between the three main accretion regimes (e.g., recurrence nova, steady burning, and giant envelope). Models that do not end in type Ia supernova events end up in the giant envelope accreting regime, given that the last thermal pulses from the donor star exhibit extreme events of mass loss. 
    
    \item The models produce bolometric luminosities consistent with optical and X-ray observations of symbiotic systems, supporting the reliability of the accretion scheme used in our simulation to match observed systems.

\end{itemize}

This study demonstrates the critical role of the adopted accretion prescription in shaping the evolution of symbiotic systems and highlights the limitations of standard BHL-based models. The results also emphasize the importance of wind accretion dynamics in reproducing observed systems, offering new insights into the progenitors and long-term evolution of symbiotic binaries. The next step will be to study the accretion scheme defined in Paper I but for the more complex interactions unveiled by eccentric cases, the inclusion of nova eruptions, and the impact of the WRLO throughout the evolution of symbiotic binaries.
\\

%\begin{acknowledgments}
\noindent We thank an anonymous referee for comments and suggestions that improved our manuscript. R.F.M. thanks UNAM DGAPA (Mexico) and SECIHTI (Mexico) for postdoc fellowships. J.A.T. and J.B.R.-G. acknowledge support from the UNAM PAPIIT project IN102324. This work has made extensive use of NASA's Astrophysics Data System. 

%\end{acknowledgments}

\appendix

\section{Avoiding the Wind Roche Lobe Overflow phase}
\label{app:WRLO}

\citet{Mohamed2012} used smoothed particle hydrodynamic numerical simulations and showed that one of the conditions for binary systems to evolve through a WRLO scheme is when $R_\mathrm{cond} \geq R_\mathrm{RL}$. Consequently, we stopped our dynamical simulations when this condition is reached by our evolving binary systems.

The condensation radius was approximated by \citep[see, e.g.,][]{Hofner2007}
\begin{equation}
    R_\mathrm{cond} = \frac{R_1}{2} \left(\frac{T_1}{T_\mathrm{cond}} \right)^{2},
\label{eq:rcond}    
\end{equation}
\noindent where $R_1$ and $T_1$ are the radius and effective temperature of the mass donor star in the system. $T_\mathrm{cond}$ is the dust condensation temperature, which in the present work will be assumed to be 1500 K \citep{Whittet2003}, appropriate for amorphous carbon. 

On the other hand, the Roche Lobe radius was calculated adopting the approximation presented in \citet{Eggleton1983}
\begin{equation}
    \frac{R_\mathrm{RL}}{a} = \frac{0.49~\delta^{2/3} }{0.6~\delta^{2/3} + \ln(1 + \delta^{1/3})},
\label{eq:RL}    
\end{equation}
\noindent with $\delta$ as the ratio of the masses of the late-type star $m_1$ over that of the accretor's $m_2$.

\begin{figure*}
\begin{center}
\includegraphics[width=\linewidth]{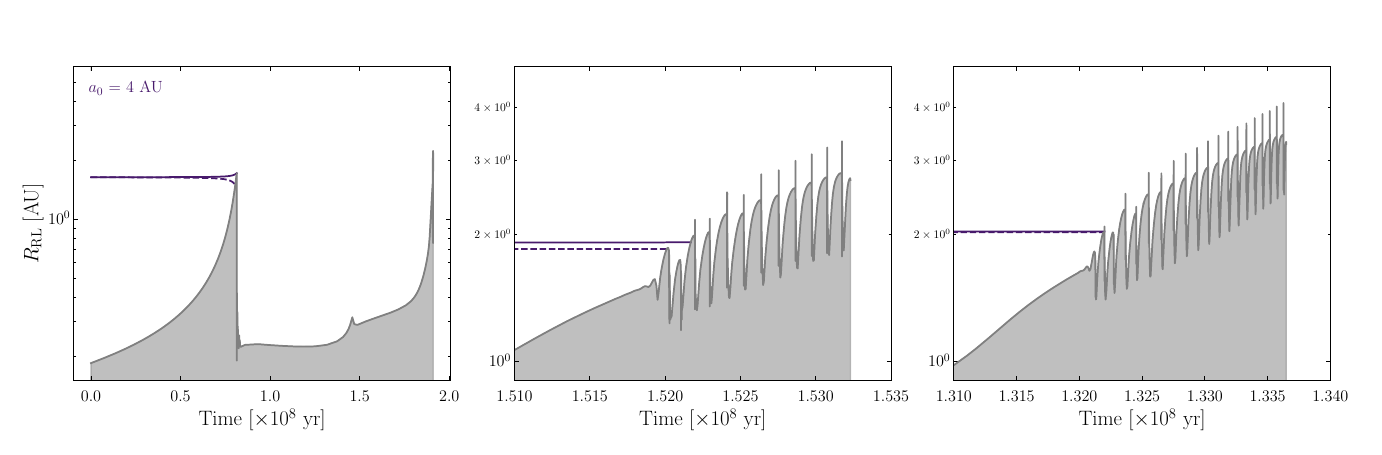}
\end{center}
\caption{Evolution of the Roche lobe radius $R_\mathrm{RL}$ of the donor star calculated by equation \ref{eq:RL} for simulations with an accretor's mass  $m_{2,0}=0.7$ M$_\odot$ in the smallest orbital configuration of our simulation's setup. The solid line depicts simulations with the new modified BHL accretion scheme while the dashed line refers to the classical BHL prescription.  The gray shaded region depicts the condensation radius calculated with equation \ref{eq:rcond} for the donor star with mass $m_{1,0}=1$ M$_\odot$ (left panel), $m_{1,0}=2$ M$_\odot$ (middle panel) and $m_{1,0}=3$ M$_\odot$ (right panel). }
\label{fig:RL}
\end{figure*}

Almost none of the modelled systems were affected by this condition. Only the most compact binary systems ($a_0 = 4$ AU) reach the $R_\mathrm{cond} = R_\mathrm{RL}$ condition and thus their calculation is stopped (see Fig. \ref{fig:RL}). However, it is worth noticing here that this condition is reached during the RGB phase in simulations with $m_{1,0}=1$ M$_\odot$ and for the other stellar models it is achieved during the evolution of the TPAGB of the mass donor star for the closest orbit. For binary separations $a_0\geq8$ AU, the system evolves without interruption until the end of the TPAGB phase or when the accretor reaches the Chandrasekhar limit (1.4 M$_\odot$).

\section{Observation Symbiotic Systems}
\label{app:symbiotics}

Table~\ref{tab:objects} lists the properties of symbiotic systems as listed in the {\it New Online Database of Symbiotic Variables}\footnote{\url{https://sirrah.troja.mff.cuni.cz/~merc/nodsv/}} \citep{Merc2019}. It lists the object name, masses of the stellar components ($m_1$ and $m_2$), the binary period ($T$), their estimated orbital separation ($a$), the orbital velocity ($v_\mathrm{o}$), the dimensionless mass ratio $q$ as defined by Eq.~(\ref{eq:q}), the dimensionless velocity parameter defined by Eq.~(\ref{eq:w}), the estimated mass accretion efficiency $\eta$ obtained from Eq.~(\ref{eq:TejedaToala}), the Roche Lobe radius calculated with Eq.~(\ref{eq:RL}) and the radius of the red giant star. We only selected symbiotic systems for which their estimated Roche Lobe radius is at least three times larger than the estimated radius of the RG star, which according to \citep{Hofner2007} is a proxy for the dust condensation radius in AGB stars.

\begin{table*}
\begin{center}
\footnotesize
\caption{Binary and orbital parameters estimated for symbiotic systems. The columns list, from left to right: mass of the primary (donor) $m_1$, mass of the secondary (accretor) $m_2$, orbital separation $a$, eccentricity $e$, orbital velocity $v_\mathrm{o}$, orbital period $T$, mass ratio $q=m_2/(m_1+m_2)$, range of wind-to-orbital velocity ratio $w = v_\mathrm{w}/v_\mathrm{o}$, calculated mass accretion efficiencies $\eta$, Roche Lobe radius $R_\mathrm{LB}$ and the red giant radius $R_\mathrm{RG}$. We assume a typical wind velocity of $v_\mathrm{w}$=12~km~s$^{-1}$. 
} 
\begin{tabular}{lccccccccccc}
\hline
Object & $m_1$      & $m_2$       & $T$  & $a$  & $e$ &$v_\mathrm{o}$ & $q$ & $w$ & $\eta$  & $R_\mathrm{RL}$ & $R_\mathrm{RG}$\\
       & [M$_\odot$]& [M$_\odot$] & [yr] &  (AU)&  & [km~s$^{-1}$]   &   &  &   & [AU] & [AU] \\
\hline
$o$ Cet &  2.00 &  0.65 & 497.90&  87.3 & 0.16  & 5.2   & 0.25 & 1.35 & $7.6\times10^{-3}$ & 42.1 & 1.9 \\
AG Dra  &  1.20 &  0.50 &  1.51 &  1.6  & 0.0   & 31.1  & 0.29 & 0.48 & $5.6\times10^{-2}$  & 0.7 & 0.16 \\
AG Peg  &  2.60 &  0.65 &  2.23 &  2.5  & 0.11  & 33.8  & 0.20 & 0.44 & $2.8\times10^{-2}$ & 1.3 & 0.4 \\
BX Mon  &  3.70 &  0.55 &  3.78 &  3.9  & 0.444 & 31.0  & 0.13 & 0.48 & $1.1\times10^{-2}$ & 2.2 & 0.7 \\
CH Cyg  &  2.20 &  0.56 & 15.58 &  8.8  & 0.122 & 16.8  & 0.20 & 0.89 & $1.2\times10^{-2}$ & 4.4 & 1.3 \\
EG And  &  1.46 &  0.40 &  1.32 &  1.5  & 0.0   & 33.5  & 0.22 & 0.45 & $3.3\times10^{-2}$ & 0.7 & 0.6 \\
ER Del  &  3.00 &  0.70 &  5.72 &  5.0  & 0.228 & 25.8  & 0.19 & 0.58 & $2.0\times10^{-2}$ & 2.5 & 0.5 \\
HD 330036 & 4.46 & 0.54 & 4.59  & 4.8   & \dots & 30.7  & 0.11 & 0.49 & $7.6\times10^{-3}$ & 2.7 & 0.1 \\
IV Vir    & 0.90 & 0.42 & 0.77  & 0.9   & 0.0   & 35.7  & 0.32 & 0.42 & $7.3\times10^{-2}$ & 0.4 & 0.1 \\
LT Del    & 1.00 & 0.57 & 1.24  & 1.4   & 0.4   & 31.9  & 0.36 & 0.47 & $8.8\times10^{-2}$ & 0.6 & 0.14 \\
PU Vul  &  1.00 &  0.50 & 13.42 &  6.5  & 0.16  & 14.4  & 0.33 & 1.04 & $2.5\times10^{-2}$ & 2.9 & 0.9 \\
R Aqr   &  1.00 &  0.70 & 42.40 &  14.6 & 0.25  & 10.2  & 0.41 & 1.47 & $1.7\times10^{-2}$ & 6.0 & 1.4 \\
St 2-22   & 2.80 & 0.80 & 2.51  & 2.9   & 0.16  & 33.6  & 0.22 & 0.45 & $3.4\times10^{-2}$ & 1.4 & 0.4 \\
TX CVn  &  3.50 &  0.40 &  0.55 &  1.1  & 0.6   & 57.4  & 0.10 & 0.26 & $8.8\times10^{-3}$ & 0.6 & 0.1 \\
V471 Per  & 2.30 & 0.80 & 17.00 & 9.4   & \dots & 16.4  & 0.26 & 0.92 & $1.9\times10^{-2}$ & 4.6 & 0.05 \\
V934 Her  & 1.60 & 1.35 & 12.02 & 7.6   & 0.354 & 18.7  & 0.46 & 0.80 & $7.7\times10^{-2}$ & 3.0 & 0.4 \\
\hline
\end{tabular}
\label{tab:objects}
\end{center}
\end{table*}


\begin{thebibliography}{99}

\bibitem[Akras et al.(2019)]{Akras2019} Akras, S., Guzman-Ramirez, L., Leal-Ferreira, M.~L., et al.\ 2019, \apjs, 240, 2, 21. doi:10.3847/1538-4365/aaf88c

\bibitem[Alcolea et al.(2023)]{Alcolea2023} Alcolea, J., Mikolajewska, J., G{\'o}mez-Garrido, M., et al.\ 2023, Highlights on Spanish Astrophysics XI, 190. %

\bibitem[Andrews et al.(2024)]{Andrews2024} Andrews, J.~J., Bavera, S.~S., Briel, M., et al.\ 2024, arXiv:2411.02376. doi:10.48550/arXiv.2411.02376

\bibitem[Beers \& Christlieb(2005)]{Beers2005} Beers, T.~C. \& Christlieb, N.\ 2005, \araa, 43, 531. doi:10.1146/annurev.astro.42.053102.134057

\bibitem[Bidelman \& Keenan(1951)]{Bidelman1951} Bidelman, W.~P. \& Keenan, P.~C.\ 1951, \apj, 114, 473. doi:10.1086/145488

\bibitem[Bloecker(1995)]{Bloecker1995} Bloecker, T.\ 1995, \aap, 297, 727. %

\bibitem[Bode(2010)]{Bode2010} Bode, M.~F.\ 2010, Astronomische Nachrichten, 331, 160. doi:10.1002/asna.200911319

\bibitem[Boffin(2015)]{Boffin2015} Boffin, H.~M.~J.\ 2015, Astrophysics and Space Science Library, 413, 153. doi:10.1007/978-3-662-44434-4\_7

\bibitem[Bondi(1952)]{Bondi1952} Bondi, H.\ 1952, \mnras, 112, 195. doi:10.1093/mnras/112.2.195

\bibitem[Bondi \& Hoyle(1944)]{Bondi1944} Bondi, H. \& Hoyle, F.\ 1944, \mnras, 104, 273. doi:10.1093/mnras/104.5.273

\bibitem[Boneva \& Yankova(2021)]{Boneva2021} Boneva, D. \& Yankova, K.\ 2021, The 20.5th Cambridge Workshop on Cool Stars, Stellar Systems, and the Sun (CS20.5), 328. doi:10.5281/zenodo.4748796

\bibitem[Boshkayev et al.(2016)]{Boshkayev2016} Boshkayev, K.~A., Rueda, J.~A., Zhami, B.~A., et al.\ 2016, International Journal of Modern Physics Conference Series, 41, 1660129. doi:10.1142/S2010194516601290

\bibitem[Camisassa et al.(2019)]{Camisassa2019} Camisassa, M.~E., Althaus, L.~G., C{\'o}rsico, A.~H., et al.\ 2019, \aap, 625, A87. doi:10.1051/0004-6361/201833822

\bibitem[Chen et al.(2018)]{Chen2018} Chen, Z., Blackman, E.~G., Nordhaus, J., et al.~2018, \mnras, 473, 747. doi:10.1093/mnras/stx2335 

\bibitem[Chomiuk et al.(2021)]{Chomiuk2021} Chomiuk, L., Metzger, B.~D., \& Shen, K.~J.\ 2021, \araa, 59, 391. doi:10.1146/annurev-astro-112420-114502

\bibitem[Curd et al.(2017)]{Curd2017} Curd, B., Gianninas, A., Bell, K.~J., et al.\ 2017, \mnras, 468, 239. doi:10.1093/mnras/stx320

\bibitem[Eggleton(1983)]{Eggleton1983} Eggleton, P.~P.\ 1983, \apj, 268, 368. doi:10.1086/160960

\bibitem[Garc{\'\i}a-Berro et al.(2012)]{GarciaBerro2012} Garc{\'\i}a-Berro, E., Lor{\'e}n-Aguilar, P., Aznar-Sigu{\'a}n, G., et al.\ 2012, \apj, 749, 25. doi:10.1088/0004-637X/749/1/25

\bibitem[Garc{\'\i}a-Berro et al.(1997)]{GarciaBerro1997} Garc{\'\i}a-Berro, E., Ritossa, C., \& Iben, I.\ 1997, \apj, 485, 765. doi:10.1086/304444

\bibitem[Gromadzki et al.(2006)]{Gromadzki2006} Gromadzki, M., et al.\ 2006, \actaa, 56, 97. doi:10.48550/arXiv.astro-ph/0604089

\bibitem[Guerrero et al.(2025)]{Guerrero2025} Guerrero, M.~A., Vasquez-Torres, D.~A., Rodr{\'\i}guez-Gonz{\'a}lez, J.~B., et al.\ 2025, \aap, 693, A203. doi:10.1051/0004-6361/202451715

\bibitem[Guerrero et al.(2024)]{Guerrero2024_XAGBs} Guerrero, M.~A., Montez, R., Ortiz, R., et al.\ 2024, \aap, 689, A62. doi:10.1051/0004-6361/202450155

\bibitem[Hachisu et al.(1996)]{Hachisu1996} Hachisu, I., Kato, M., \& Nomoto, K.\ 1996, \apjl, 470, L97. doi:10.1086/310303

\bibitem[Hansen et al.(2016)]{Hansen2016} Hansen, T.~T., Andersen, J., Nordstr{\"o}m, B., et al.\ 2016, \aap, 588, A3. doi:10.1051/0004-6361/201527409 

\bibitem[Hermes et al.(2013)]{Hermes2013} Hermes, J.~J., Kepler, S.~O., Castanheira, B.~G., et al.\ 2013, \apjl, 771, L2. doi:10.1088/2041-8205/771/1/L2

\bibitem[Hillman \& Kashi(2021)]{HillmanKashi2021} Hillman, Y. \& Kashi, A.\ 2021, \mnras, 501, 201. doi:10.1093/mnras/staa3600

\bibitem[H{\"o}fner(2007)]{Hofner2007} H{\"o}fner, S.\ 2007, 378, 145. doi:10.48550/arXiv.astro-ph/0702444

\bibitem[Hoyle \& Lyttleton(1939)]{Hoyle1939} Hoyle, F. \& Lyttleton, R.~A.\ 1939, Proceedings of the Cambridge Philosophical Society, 35, 405. doi:10.1017/S0305004100021150

\bibitem[Huarte-Espinosa et al.(2013)]{HE2013} Huarte-Espinosa, M., Carroll-Nellenback, J., Nordhaus, J., et al.\ 2013, \mnras, 433, 295. doi:10.1093/mnras/stt725

\bibitem[Hurley et al.(2002)]{Hurley2002} Hurley, J.~R., Tout, C.~A., \& Pols, O.~R.\ 2002, \mnras, 329, 897. doi:10.1046/j.1365-8711.2002.05038.x

\bibitem[Izzard et al.(2009)]{Izzard2009} Izzard, R.~G., Glebbeek, E., Stancliffe, R.~J., et al.\ 2009, \aap, 508, 1359. doi:10.1051/0004-6361/200912827

\bibitem[Jermyn et al.(2023)]{Jermyn2023} Jermyn, A.~S., Bauer, E.~B., Schwab, J., et al.,\ 2023, \apjs, 265, 15. doi:10.3847/1538-4365/acae8d

\bibitem[Kato et al.(2024)]{Kato2024} Kato, M., Saio, H., \& Hachisu, I.\ 2024, \pasj, 76, 4, 666. doi:10.1093/pasj/psae038

\bibitem[Kato et al.(2022)]{Kato2022} Kato, M., Saio, H., \& Hachisu, I.\ 2022, \pasj, 74, 5, 1005. doi:10.1093/pasj/psac051

\bibitem[Kato et al.(2017)]{Kato2017} Kato, M., Saio, H., \& Hachisu, I.\ 2017, \apj, 838, 2, 153. doi:10.3847/1538-4357/838/2/153

\bibitem[Karinkuzhi et al.(2024)]{Karinkuzhi2024} Karinkuzhi, D., Mukhopadhyay, B., Wickramasinghe, D., et al.\ 2024, \mnras, 529, 4577. doi:10.1093/mnras/stae829

\bibitem[Kenyon \& Truran(1983)]{Kenyon1983} Kenyon, S.~J. \& Truran, J.~W.\ 1983, \apj, 273, 280. doi:10.1086/161367

\bibitem[Leedjarv et al.(1994)]{Leedjarv1994} Leedjarv, L., Mikolajewski, M., \& Tomov, T.\ 1994, \aap, 287, 543. %

\bibitem[Li et al.(2023)]{Li2023} Li, Z., Chen, X., Ge, H., et al.\ 2023, \aap, 669, A82. doi:10.1051/0004-6361/202243893

\bibitem[Lima et al.(2024)]{Lima2024} Lima, I.~J., Luna, G.~J.~M., Mukai, K., et al.\ 2024, \aap, 689, A86. doi:10.1051/0004-6361/202449913

\bibitem[Liu et al.(2017)]{Liu2017} Liu, Z.-W., Stancliffe, R.~J., Abate, C., et al.\ 2017, \apj, 846, 117. doi:10.3847/1538-4357/aa8622

\bibitem[Luna et al.(2018)]{Luna2018} Luna, G.~J.~M., Mukai, K., Sokoloski, J.~L., et al.\ 2018, \aap, 616, A53. doi:10.1051/0004-6361/201832592 

\bibitem[Luna et al.(2013)]{Luna2013} Luna, G.~J.~M., Sokoloski, J.~L., Mukai, K., et al.\ 2013, \aap, 559, A6. doi:10.1051/0004-6361/201220792

\bibitem[Malfait et al.(2024)]{Malfait2024} Malfait, J., Siess, L., Esseldeurs, M., et al.\ 2024, \aap, 691, A84. doi:10.1051/0004-6361/202450338

\bibitem[Marchev \& Zamanov(2024)]{Marchev2024} Marchev, V.~D. \& Zamanov, R.~K.\ 2024, Bulgarian Astronomical Journal, 40, 85. doi:10.48550/arXiv.2312.04208

\bibitem[Merc(2025)]{Merc2025} Merc, J.\ 2025,  arXiv:2504.16825  doi:10.48550/arXiv.2504.16825

\bibitem[Merc et al.(2024)]{Merc2024} Merc, J., Beck, P.~G., Mathur, S., et al.\ 2024, \aap, 683, A84. doi:10.1051/0004-6361/202348116

\bibitem[Merc et al.(2019)]{Merc2019} Merc, J., G{\'a}lis, R., \& Wolf, M.\ 2019, Research Notes of the American Astronomical Society, 3, 28. doi:10.3847/2515-5172/ab0429

\bibitem[Mohamed \& Podsiadlowski(2012)]{Mohamed2012} Mohamed, S. \& Podsiadlowski, P.\ 2012, Baltic Astronomy, 21, 88. doi:10.1515/astro-2017-0362

\bibitem[Mohamed \& Podsiadlowski(2007)]{Mohamed2007} Mohamed, S. \& Podsiadlowski, P.\ 2007, 15th European Workshop on White Dwarfs, 372, 397. 

\bibitem[Mukai(2017)]{Mukai2017} Mukai, K.\ 2017, \pasp, 129, 062001. doi:10.1088/1538-3873/aa6736

\bibitem[Nagae et al.(2004)]{Nagae2004} Nagae, T., Oka, K., Matsuda, T., et al.\ 2004, \aap, 419, 335. doi:10.1051/0004-6361:20040070

\bibitem[Nomoto et al.(2007)]{Nomoto2007} Nomoto, K., Saio, H., Kato, M., et al.\ 2007, \apj, 663, 1269. doi:10.1086/518465

\bibitem[Osborn et al.(2025)]{Osborn2025} Osborn, Z., Karakas, A., Kemp, A., et al.\ 2025, \pasa, 42, e020. doi:10.1017/pasa.2024.124

\bibitem[Pasquini et al.(2023)]{Pasquini2023} Pasquini, L., Pala, A.~F., Salaris, M., et al.\ 2023, \mnras, 522, 3710. doi:10.1093/mnras/stad1252

\bibitem[Paxton et al.(2019)]{Paxton2019} Paxton, B., Smolec, R., Schwab, J., et al.\ 2019, \apjs, 243, 10. doi:10.3847/1538-4365/ab2241 

\bibitem[Paxton et al.(2018)]{Paxton2018} Paxton, B., Schwab, J., Bauer, E.~B., et al.\ 2018, \apjs, 234, 34. doi:10.3847/1538-4365/aaa5a8

\bibitem[Paxton et al.(2015)]{Paxton2015} Paxton, B., Marchant, P., Schwab, J., et al.\ 2015, \apjs, 220, 15. doi:10.1088/0067-0049/220/1/15

\bibitem[Paxton et al.(2013)]{Paxton2013} Paxton, B., Cantiello, M., Arras, P., et al.\ 2013, \apjs, 208, 4. doi:10.1088/0067-0049/208/1/4

\bibitem[Paxton et al.(2011)]{Paxton2011} Paxton, B., Bildsten, L., Dotter, A., et al.\ 2011, \apjs, 192, 3. doi:10.1088/0067-0049/192/1/3

\bibitem[Pringle(1981)]{Pringle1981} Pringle, J.~E.\ 1981, \araa, 19, 137. doi:10.1146/annurev.aa.19.090181.001033

\bibitem[Pujol et al.(2023)]{Pujol2023} Pujol, A., Luna, G.~J.~M., Mukai, K., et al.\ 2023, \aap, 670, A32. doi:10.1051/0004-6361/202244967

\bibitem[Ramstedt et al.(2020)]{Ramstedt2020} Ramstedt, S., Vlemmings, W.~H.~T., Doan, L., et al.\ 2020, \aap, 640, A133. doi:10.1051/0004-6361/201936874

\bibitem[Ramstedt et al.(2018)]{Ramstedt2018} Ramstedt, S., Mohamed, S., Olander, T., et al.\ 2018, \aap, 616, A61. doi:10.1051/0004-6361/201833394

\bibitem[Reimers(1975)]{Reimers1975} Reimers, D.\ 1975, Memoires of the Societe Royale des Sciences de Liege, 8, 369. %

\bibitem[Rein \& Spiegel(2015)]{Rein2015} Rein, H. \& Spiegel, D.~S.\ 2015, \mnras, 446, 1424. doi:10.1093/mnras/stu2164

\bibitem[Rein \& Liu(2012)]{Rein2012} Rein, H. \& Liu, S.-F.\ 2012, \aap, 537, A128. doi:10.1051/0004-6361/201118085

\bibitem[Richards et al.(2024)]{Richards2024} Richards, S., Eldridge, J., Ghodla, S., et al.\ 2024, arXiv:2411.03000. doi:10.48550/arXiv.2411.03000

\bibitem[Robinson et al.(1994)]{Robinson1994} Robinson, K., Bode, M.~F., Skopal, A., et al.\ 1994, \mnras, 269, 1. doi:10.1093/mnras/269.1.1

\bibitem[Saladino \& Pols(2019)]{SaladinoPols2019} Saladino, M.~I. \& Pols, O.~R.\ 2019, \aap, 629, A103. doi:10.1051/0004-6361/201935625

\bibitem[Saladino et al.(2019)]{Saladino2019} Saladino, M.~I., Pols, O.~R., \& Abate, C.\ 2019, \aap, 626, A68. doi:10.1051/0004-6361/201834598

\bibitem[Schaefer(2023)]{Schaefer2023} Schaefer, B.~E.\ 2023, Journal for the History of Astronomy, 54, 436. doi:10.1177/00218286231200492

\bibitem[Shakura \& Sunyaev(1973)]{Shakura1973} Shakura, N.~I. \& Sunyaev, R.~A.\ 1973, \aap, 24, 337. %

\bibitem[Shara et al.(2018)]{Shara2018} Shara, M.~M., Prialnik, D., Hillman, Y., et al.\ 2018, \apj, 860, 110. doi:10.3847/1538-4357/aabfbd

\bibitem[Shen \& Bildsten(2007)]{Shen2007} Shen, K.~J. \& Bildsten, L.\ 2007, \apj, 660, 1444. doi:10.1086/513457

\bibitem[Sokoloski et al.(2001)]{Sokoloski2001} Sokoloski, J.~L., Bildsten, L., \& Ho, W.~C.~G.\ 2001, \mnras, 326, 553. doi:10.1046/j.1365-8711.2001.04582.x

\bibitem[Stanishev et al.(2004)]{Stanishev2004} Stanishev, V., Zamanov, R., Tomov, N., et al.\ 2004, \aap, 415, 609. doi:10.1051/0004-6361:20034623

\bibitem[Starrfield \& Sparks(1987)]{Starrfield1987} Starrfield, S. \& Sparks, W.~M.\ 1987, \apss, 131, 379. doi:10.1007/BF00668117

%\bibitem[Tamayo et al.(2019)]{Tamayo2019} Tamayo, D., Rein, H., Shi, P., Hernandez, D.~M. \ 2019, \mnras, 491, 2885. %doi.org/10.1093/mnras/stz2870

\bibitem[Tejeda \& Toal{\'a}(2025)]{TejedaToala2025} Tejeda, E. \& Toal{\'a}, J.~A.\ 2025, \apj, 980, 2, 226. doi:10.3847/1538-4357/ada953

\bibitem[Theuns et al.(1996)]{Theuns1996} Theuns, T., Boffin, H.~M.~J., \& Jorissen, A.\ 1996, \mnras, 280, 1264. doi:10.1093/mnras/280.4.1264

\bibitem[Toal{\'a} et al.(2024)]{Toala2024} Toal{\'a}, J.~A., Gonz{\'a}lez-Mart{\'\i}n, O., Sacchi, A., et al.\ 2024, \mnras, 532, 1421. doi:10.1093/mnras/stae1579

\bibitem[Toal{\'a} et al.(2023)]{Toala2023} Toal{\'a}, J.~A., Gonz{\'a}lez-Mart{\'\i}n, O., Karovska, M., et al.\ 2023, \mnras, 522, 6102. doi:10.1093/mnras/stad1401

\bibitem[Vathachira et al.(2024)]{Vathachira2024} Vathachira, I.~B., Hillman, Y., \& Kashi, A.\ 2024, \mnras, 527, 4806. doi:10.1093/mnras/stad3507 

\bibitem[Vasquez-Torres et al.(2024)]{VT2024} Vasquez-Torres, D.~A., Toal{\'a}, J.~A., Sacchi, A., et al.\ 2024, \mnras, 535, 2724. doi:10.1093/mnras/stae2538

\bibitem[Vennes \& Kawka(2008)]{Vennes2008} Vennes, S. \& Kawka, A.\ 2008, \mnras, 389, 1367. doi:10.1111/j.1365-2966.2008.13652.x

\bibitem[Verbena et al.(2011)]{Verbena2011} Verbena, J.~L., Schr{\"o}der, K.-P., \& Wachter, A.\ 2011, \mnras, 415, 2270. doi:10.1111/j.1365-2966.2011.18859.x

\bibitem[Whittet(2003)]{Whittet2003} Whittet, D.~C.~B.\ 2003, Dust in the galactic environment, Dust in the galactic environment. %

\bibitem[Wolf et al.(2013)]{Wolf2013} Wolf, W.~M., Bildsten, L., Brooks, J., et al.\ 2013, \apj, 777, 136. doi:10.1088/0004-637X/777/2/136

\bibitem[Zamanov et al.(2024)]{Zamanov2024} Zamanov, R.~K., Stoyanov, K.~A., Marchev, V., et al.\ 2024, Astronomische Nachrichten, 345, e20240036. doi:10.1002/asna.20240036

\bibitem[Zhekov \& Tomov(2019)]{Zhekov2019} Zhekov, S.~A. \& Tomov, T.~V.\ 2019, \mnras, 489, 2930. doi:10.1093/mnras/stz2329

\end{thebibliography}
\end{document}